\documentclass[a4paper,fleqn,usenatbib]{mnras}

\usepackage{newtxtext,newtxmath}

\usepackage[T1]{fontenc}
\usepackage{ae,aecompl}

\usepackage{graphicx} 
\usepackage{amsmath} 
\usepackage{amssymb} 

\usepackage{amssymb}
\usepackage{amsmath}
\usepackage{lscape}
\usepackage{morefloats}
\usepackage{bigints}
\usepackage[font=small,labelfont=bf]{caption}
\usepackage{graphicx}
\usepackage{mathtools}
\usepackage[titletoc,toc,title]{appendix}

\usepackage{float}
\usepackage{wrapfig}
\usepackage{lscape}
\usepackage{rotating}
\usepackage{breqn}
\usepackage{epstopdf}
\usepackage[bottom]{footmisc}
\usepackage[normalem]{ulem}

\title[Baryon impacts on dark energy]{The impact of baryons on the sensitivity of dark energy measurements}

\author[D.N. Copeland, A.N. Taylor, Alex Hall]{
David Copeland,$^{1}$\thanks{E-mail: dcope@roe.ac.uk}
Andy Taylor,$^{1}$
Alex Hall, $^{1}$
\\
$^{1}$Institute for Astronomy, University of Edinburgh, Royal Observatory, Blackford Hill, Edinburgh EH9 3HJ
}

\date{Accepted XXX. Received YYY; in original form ZZZ}

\pubyear{2017}

\bibpunct[, ]{(}{)}{;}{a}{}{,}
\AtBeginShipout{%
  \ifnum\value{page}>1 %
    \typeout{* Additional boxing of page `\thepage'}%
    \setbox\AtBeginShipoutBox=\hbox{\copy\AtBeginShipoutBox}%
  \fi
}

\begin{document}
\label{firstpage}
\pagerange{\pageref{firstpage}--\pageref{lastpage}}
\maketitle

\defcitealias{Mead15}{M15}

\begin{abstract}
Baryonic effects on large-scale structure, if not accounted for, can significantly bias dark energy constraints. As the detailed physics of the baryons is not yet well-understood, correcting for baryon effects introduces additional parameters which must be marginalized over, increasing the uncertainties on the inferred cosmological parameters. Forthcoming weak lensing surveys are aiming for percent-level precision on the dark energy equation of state, so the problem must be thoroughly examined. We use a halo model with analytic modifications which capture the impact of adiabatic contraction of baryons and feedback on the matter power spectrum, and generalize the Navarro-Frenk-White profile to account for a possible inner core. A Fisher analysis predicts degradations of 40\% in the $w_0$-$w_a$ Figure of Merit for a Euclid-like survey, and up to 80\% for other cosmological parameters. We forecast potential inner core constraints of a few $\mathrm{kpc}$, while for a fixed inner core, adiabatic concentration and feedback parameters are constrained to a few percent. We explore the scales where baryons and dark energy contribute most to the Fisher information, finding that probing to increasingly non-linear scales does little to reduce degradation. Including external baryon information improves our forecasts, but limiting degradation to 1\% requires strong priors. Adding \textit{Planck} cosmic microwave background priors improves the Figure of Merit by a factor of 2.7 and almost completely recovers the individual marginalized errors on $w_0$ and $w_a$. We also quantify the calibration of baryon modelling required to reduce biases of dark energy forecasts to acceptable levels for forthcoming lensing surveys.
\end{abstract}


\section{Introduction}
Understanding the present accelerating expansion of the Universe has been one of the primary goals of cosmology since the $\Lambda$CDM model was favoured by observations of type Ia supernovae \citep{jr:riess98, Perlmutter99}. An enduring obstacle is the discrepancy between the observed value of the cosmological constant, $\Lambda$, and the far larger quantum field theory predictions for the gravitation of vacuum energy \citep[e.g.,][]{Weinberg89, Burgess13}. Without solving this problem, alternative acceleration mechanisms fall into two broad categories: that some undetected `dark energy' permeating space is responsible \citep[see reviews by][]{Copeland06, Frieman08}; or that the effect emerges from modifications to general relativity \citep[e.g.,][]{Clifton11}. 

Large-scale structure is a useful probe for constraining this model space, as dark energy suppresses the growth of the dark matter distribution relative to the Einstein-de Sitter case. However, it becomes increasingly challenging to model accurately the collapse and growth of density fluctuations once they give rise to the web of haloes and connecting filaments characterizing the non-linear regime. A valuable tool for exploring these scales is tomographic weak gravitational lensing \citep{Bartelmann&Schneider01}, which uses multiple photometric redshift bins to disentangle information about the sensitivity of the expansion history and structure growth to dark energy \citep{Albrecht06, Peacock06}. Indeed, forthcoming Stage IV surveys like Euclid \citep{Laurejis11} or LSST \citep{LSST09} anticipate percent-level precision on measurements of dark energy parameters. These are defined by the time-varying dark energy equation-of-state \citep{Chevallier01, Linder03}, 
\begin{equation}
w\left(a\right)=w_0+w_a\left(1-a\right),
\end{equation}
where $w\equiv P/\rho$ is the ratio of dark energy pressure to energy density. The present value of $w$ is given by $w_0$ while $w_a$ specifies its rate of change, $w_a\equiv -\mathrm{d}w/\mathrm{d}a$, with respect to the scale factor, $a$, parameterizing the cosmic expansion.

A serious risk to this aim is that matter is also substantially redistributed on halo scales by baryonic processes, such as radiative cooling or energetic events like active galactic nuclei (AGN) and supernovae \citep[see e.g.,][]{Rudd08, vandaalen14}. These processes are not well-understood so they potentially introduce enough uncertainty into parameter forecasts to significantly degrade dark energy constraints. High-resolution hydrodynamic simulations like the OverWhelmingly Large Simulations project \citep[OWLS;][]{Schaye10} have been used to show that transactions of thermal energy to the local environment by AGN feedback can reduce the baryon fraction by several factors in the inner regions of haloes and bloat the matter distribution \citep[e.g.,][]{Duffy10}. This effect acts in opposition to the adiabatic contraction that occurs from dark matter infall into a potential well that has been strengthened by cooling gas clustering on small scales \citep{Gnedin04, Jing06, Rudd08, Duffy10}. The net impact on the matter power spectrum is suppression by up to 30\% in the non-linear regime, which is overtaken beyond $k\sim10\, h\, \mathrm{Mpc}^{-1}$ by enhancements to power from halo concentrations increasing due to radiative cooling  \citep[e.g.,][]{Semboloni11}. 

This has lead to several recalibrations of the halo model via simulations to account for baryons \citep[see e.g.,][]{Semboloni11,Zentner12,Semboloni13,Mohammed14,Mead15,Schneider&Teyssier15}. The sensitivity of the weak lensing projection of the matter power spectrum to baryon and dark energy parameters can then be examined. Different approaches for implementing baryons have generated a considerable range of results for the impact on dark energy error forecasts. Among the most alarming are those of \citet{Zentner12} who find, by modifying the halo concentration relation, that 1-$\sigma$ errors for $w_0$ and $w_a$ increase by $\sim50$\%. These degradations would compound to severely reduce the Figure of Merit, given by
\begin{equation}
\mathrm{FOM}=\frac{1}{\sqrt{\left(\sigma_{w_0}\sigma_{w_a}\right)^2-\sigma_{w_0 w_a}^2}},
\end{equation}
that constrains the $w_0$-$w_a$ parameter space. The impact of baryons also biases estimates of the most likely parameter values. \citet{Semboloni11} determine a bias as high as 40\% in predictions of $w_0$ when neglecting for baryons which reduces to $\sim 10$\% when accounting for feedback by fitting mass fractions for separate profiles for dark matter, gas and stars. \citet{Semboloni13} argue that this overcomes a shortcoming in the modelling of \citet{Zentner12} that neglects differences between the distributions of dark matter and hot gas. \citet{Mohammed14} adopt a similar approach, by modelling stellar contributions with a central galaxy, introducing a hot plasma in hydrostatic equilibrium and accounting for the baryon-induced adiabatic contraction of dark matter due to cooling. They find a degradation of $\sim 10$\% and $\sim 30$\% to the forecasted errors on $w_0$ and $w_a$ respectively. 

The approach of these works \citep[see also][]{Fedeli14, Fedelietal14} of fitting for stellar and gas physics within the halo is different to that advocated by \citet{Mead15} (hereafter \citetalias{Mead15}). Their corrections to the halo model \citep[HMCODE; extended in][]{Mead16} are designed specifically to calibrate the power spectrum accurately for the non-linear regime. This requires empirically motivated baryon modifications to the internal halo structure relations that can be directly associated with adiabatic contraction and feedback. As the power spectrum is the statistic underpinning our study of forecast degradations, we adopt the model of \citetalias{Mead15} as the most suitable for our purposes. 

This paper extends HMCODE to include a generic treatment of inner halo cores. We are motivated here by e.g., \citet{Martizzi12} and \citet{Governato12} who show that baryons can produce inner cores of the order of $10\, \mathrm{kpc}$. Possible mechanisms range from dynamical friction effects in black hole orbit decays to AGN feedback removing dark matter from central regions by disturbing the gravitational potential. Alternatively, axions could be responsible in the form of solitons \citep{Marsh15}. Whatever the underlying physics, our version of the baryon-halo model is the basis for a more comprehensive and robust analysis of the baryon impact on dark energy constraints than has been seen previously. Our framework also allows us to explore how well the baryon parameters used in our modified HMCODE could be constrained by Stage IV surveys, given the uncertainty in cosmology. It should be stressed that this does not amount to direct constraints on baryonic phenomena, but on the approximate redistribution of matter in haloes. This could provide opportunities for informing future hydrodynamic simulations, and insight into the possible nature of cores. 

To understand the scope of baryon impacts it is important to analyse the sensitivity of information from baryons and dark energy at different lensing scales. This goes beyond examining the effect of increasing the scale limit, $\ell_{\rm{max}}$, of an analysis \citep[e.g.,][]{Semboloni11, Zentner12, Mohammed14}. Weak lensing power responses to varying $w_0$ and $w_a$ exhibit subtle scale-dependencies due to competing influences on the growth of structure and the geometry governing distances. Understanding how the $w_0$-$w_a$ degeneracy is broken in this interpretation, and how this is complicated by the inclusion of baryons, is essential for informing a mitigation strategy. Our prescription is then based on improvements offered by changing $\ell_{\rm{max}}$, incorporating external baryon information, and adding \textit{Planck} CMB priors. The last step makes use of strong constraints that have not been available for previous baryon impact studies. At the same time the advent of next generation surveys is fast approaching so now is an optimal moment to revisit the issue. The accuracy of power spectra provided by HMCODE and the scope of effects available through our inclusion of inner cores make this work uniquely placed to assess baryon degradation and target how to mitigate it. 

The paper is structured as follows. In \S~\ref{sec:BHM} we explain the baryon-halo model and our modifications for inner cores. In \S~\ref{sec:impact} we present a Fisher analysis to evaluate baryon degradation on forecasts of dark energy errors. Though we focus on results for a Euclid-like survey, our methods are applicable to other Stage IV, space-based surveys such as LSST. \S~\ref{sec:mitigation} then focuses on what the scale-dependence of dark energy and baryon Fisher information implies for mitigation strategies based on changing $\ell_{\rm{max}}$. We evaluate how much additional baryon or cosmology information is required from independent sources to make substantial improvements to the dark energy FOM. Finally, in \S~\ref{sec:modbias} we discuss the prevalence of model bias in our approach, before concluding in \S~\ref{sec:conc}.

\vspace{-6.2mm}

\section{Baryon-Halo Model}
\label{sec:BHM}
\subsection{The Halo Model}
The halo model is a powerful tool for describing the non-linear clustering of matter. It allows the large-scale galaxy distribution to be well-approximated by treating the halo occupation number for galaxies as a function of halo mass, and positioning within each halo a central galaxy around which other galaxies trace the halo profile as satellites \citep{Seljak00,Peacock00}.

An evolving, comoving spatial perturbation, $\delta\left(\bmath{x},t\right)$, about the mean cosmological matter density, $\bar{\rho}\left(t\right)$, is defined such that the matter density field is written as
\begin{equation}
\rho\left(\bmath{x},t\right) = \bar{\rho}\left(t\right)\left[1+\delta\left(\bmath{x},t\right)\right].
\end{equation} 
The matter power spectrum, $P\left(k\right)$, is then defined as
\begin{equation}
\left\langle \delta\left(\bmath{k}\right)\delta^{\star}\left(\bmath{k}^{\prime}\right) \right\rangle = \left(2\pi\right)^3 \delta_D\left(\bmath{k}-\bmath{k^{\prime}}\right) P\left(k\right).
\label{eq:powerspecdef}
\end{equation} 
More convenient for our purposes is the dimensionless form, 
\begin{equation}
\Delta^2\left(k\right)  =  \frac{k^3}{2\pi^2}P\left(k\right),
\end{equation}  
which is equivalent to the fractional contribution to the variance of the matter distribution per logarithmic interval of $k$, 
\begin{equation}
\sigma^2\left(R\right) = \int_0^{\infty} \mathrm{d}\ln k\; \Delta^2\left(k\right) W^2\left(k R\right), 
\end{equation}
where the field is smoothed over some scale $R$ using the window function of a spherical top-hat profile,
\begin{equation}
W\left(x\right) = \frac{3}{x^3}\left(\sin x - x\cos x\right). 
\end{equation}

In the halo model a distribution of spherically-collapsed halo structures randomly populate the linear density field, allowing for the effective separation of power into two distinct source terms,
\begin{equation}
\Delta^2\left(k\right) = \Delta_{1h}^2\left(k\right) + \Delta^2_{2h}\left(k\right).
\end{equation} 
The 2-halo term, $\Delta_{2h}^2\left(k\right)$, describes the correlations in the distribution of haloes themselves. As this occurs on large scales, an acceptable approximation is to equate this term to the linear matter power spectrum,
\begin{equation}
\Delta_{2h}^2\left(k\right) = \Delta_{\mathrm{lin}}^2\left(k\right).
\end{equation}
By contrast, the 1-halo term, $\Delta_{1h}^2\left(k\right)$, represents the internal halo structure on small scales. Computing this statistic requires averaging the self-convolutions of haloes over the full range of halo masses, weighted by the total number density of pairs of haloes of mass $M$. In Fourier space these convolutions become simple multiplication operations, leading to the integral
\begin{equation}
\Delta_{1h}^2\left(k\right) = \frac{k^3}{2\pi^2}\int_0^{\infty}\mathrm{d}M\, \frac{M^2 n\left(M\right)}{\bar{\rho}^2} u^2\left(k\mathopen{|}\mathclose M\right)
\label{eq:1halo} 
\end{equation}
where $n\left(M\right)$ is the comoving number density of halos per mass interval $\mathrm{d}M$, known as the halo mass function, and $u\left(k\mathopen{|}\mathclose M\right)$ is the halo density profile in Fourier space. Assuming spherical symmetry, this can be written as the transform,
\begin{equation}
u\left(k\mathopen{|}\mathclose M\right) = \frac{4\pi}{M}\int_0^{r_v}r^2\mathrm{d}r\, \frac{\sin\left(kr\right)}{kr}\rho\left(r, M\right),
\end{equation}
with the prefactor normalizing the profile by halo mass. The virial equilibrium of energy exchange between gravitationally interacting matter shells is a natural threshold at which to truncate the profile. A halo is therefore characterized by its virial density, $\Delta_v$, and radius, $r_v$, which are related by 
\begin{equation}
r_v=\left(\frac{3M}{4\pi\bar{\rho}\Delta_v}\right)^{\frac{1}{3}}.
\end{equation}
Spherical collapse calculations in the relevant cosmology inform $\Delta_v$ so the virial radius is fixed for a given halo mass. 

The form of the density profile is typically a matter of fitting to simulations of collisionless dark matter particles. The most common is the NFW profile \citep*{NFW97},
\begin{equation}
\rho\left(r,M\right) = \frac{\rho_s}{\left(\frac{r}{r_s}\right)\left[1+\left(\frac{r}{r_s}\right)\right]^2}.
\label{eq:NFW}
\end{equation}
The scale radius, $r_s$, defines a break scale between the linear and cubic declines of density in the inner and outer regions of the halo respectively. This scale also dictates the normalization factor, $\rho_s$. 
\subsection{Parameterizing Baryon Physics}
To incorporate baryons into the halo model we adopt and extend the treatment of \citetalias{Mead15}. Three general but distinct baryonic effects are parameterized: large-scale adiabatic contraction caused by radiative cooling; high-impact energy transfer from localized sources; and the formation of inner halo cores with radius $r_b$ due to small-scale physics. \citetalias{Mead15} capture the first two by varying internal halo structure relations through their parameters, $A_B$ (referred to as $A$ in \citetalias{Mead15}) and $\eta_0$. An inner core is discussed in \S~\ref{subsec:baryoncores}. 

Multiple sources of baryon physics are implemented in OWLS. In \citetalias{Mead15}, fits are made to a dark-matter only model as a fiducial model and then to three baryonic models. These include prescriptions for radiative cooling, different strengths of supernovae and AGN feedback, and various stellar processes. A fit is also made to power spectra generated by an `emulator' code (COSMIC EMU) for the high resolution N-body simulations from the Coyote Universe project \citep{Heitmann09, Heitmann10, Lawrence10, Heitmann14}. Here HMCODE achieves $\simeq 5$ percent accuracy for scales $k \leq 10\,h\,\mathrm{Mpc}^{-1}$ and redshifts $z \leq 2$, improving by several factors over HALOFIT at non-linear scales. Therefore, we take the baryon parameter values that best fit COSMIC EMU as our fiducial values in this work.

It is worth noting that \citetalias{Mead15} identify a degeneracy between $A_B$ and $\eta_0$ from fitting to multiple OWLS simulations. A likelihood analysis by \citet{Hildebrandt17} exploits this by fixing $\eta_0=1.03-0.11A_B$, where $A_B$ becomes the single free baryon parameter. We retain both parameters as they allow us to characterize multiple baryon effects and to explore how well surveys could constrain these particular phenomena, given the uncertainty in cosmology.

Throughout this paper, we use the parameter values corresponding to the base $\Lambda$CDM \textit{Planck} TT,TE,EE+lowP likelihood \citep[see Table 4 in][]{Planck15} for our fiducial cosmology, and choose $w_{0,\rm{fid}}=-1$, $w_{a,\rm{fid}}=0$. 

\subsubsection{Adiabatic Contraction}
Adiabatic contraction is the most straightforward baryon effect to model. Clustering of baryonic matter due to radiative cooling induces the gravitational infall of dark matter, so the total matter distribution undergoes contraction. Simulations have shown that the impact is at several percent for non-linear clustering \citep{Duffy10, Gnedin11}. 

\begin{figure}
\includegraphics[width=\columnwidth]{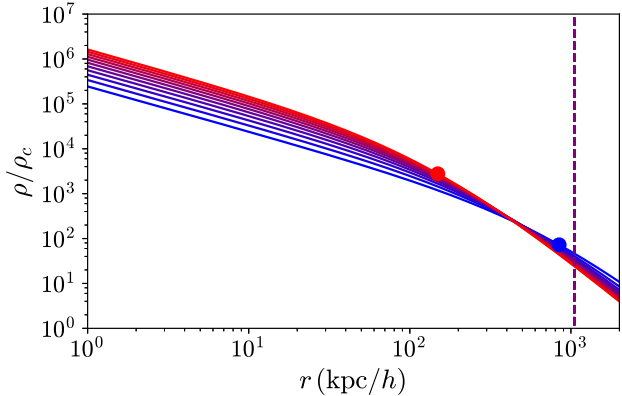}
\caption{NFW halo density profiles (in units of the critical density, $\rho_c$) for a halo of mass $M=3\times 10^{14}\, M_{\odot}$ at $z=0$ with different amplitudes, $A_B$, of the concentration factor. Blue (red) curves correspond to the lowest (highest) values in the range $0.3\,A_{B}\leq A_{B,\rm{fid}} \leq 1.7\,A_{B}$. Blue (red) dots indicate the scale radius, $r_s=845\, \left(149\right)\, h^{-1} {\rm{kpc}} $, corresponding to the lowest (highest) $A_B$ values. The purple dashed line marks the virial radius, $r_v=1050\,h^{-1} {\rm{kpc}}$, at which the profile must be truncated.}
\label{fig:density_nfw_realspace_ab} 
\end{figure}

In \citetalias{Mead15} this is captured by modifying the concentration factor, $c\left(M,z\right)$, which relates the scale and virial radii via $r_s=c r_v$. The amplitude, $A_B$, in the concentration factor
\begin{equation}
c\left(M,z\right)=A_B\frac{1+z_f}{1+z}
\end{equation}  
is allowed to vary around a fiducial value, $A_{B,\rm{fid}}=3.13$. This was chosen because \citetalias{Mead15} found it produced the best fit to COSMIC EMU power spectra. Fits by \citetalias{Mead15} to the different OWLS simulations satisfy the range $2 < A_B < 4$, which could be used to inform a prior. The dependence on halo mass enters the above expression via $z_f$, the formation redshift at which a fraction $f=0.01$ of the total matter in a density fluctuation has collapsed. Figure~\ref{fig:density_nfw_realspace_ab} shows the impact of varying the concentration amplitude on the NFW density profile of a halo with fixed virial radius and mass. Adiabatic contraction then manifests through a reduced scale radius. This corresponds to suppressing the halo density on large scales, $r\gtrsim{10^2\,h^{-1} \mathrm{kpc}}$, while enhancing it at smaller scales, $r\lesssim{10\,h^{-1} \mathrm{kpc}}$.    
\\

\vspace{-2.2mm}

\subsubsection{Baryonic Feedback}
\begin{figure}
\includegraphics[width=\columnwidth]{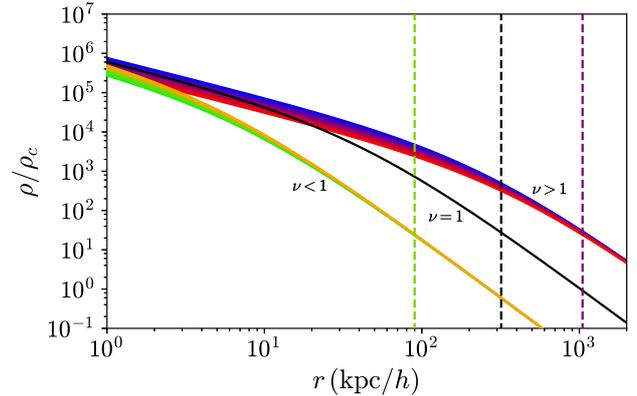}
\caption{NFW halo density profiles (in units of the critical density, $\rho_c$) for haloes of mass $M=2\times 10^{11}\, M_{\odot}$ ($\nu<1$) and $M=3\times 10^{14}\, M_{\odot}$ ($\nu>1$) at $z=0$. Green (orange) curves for the low mass case and blue (red) curves for the high mass case correspond to the lowest (highest) values of the feedback parameter in the range $0.5\,\eta_{0,\rm{fid}}\leq \eta_{0} \leq 1.5\,\eta_{0,\rm{fid}}$. The green (purple) dashed lines mark the virial radius $r_v=89.9\, \left(1050\right)\,h^{-1} {\rm{kpc}}$ for the low (high) mass halo. We also include a reference profile (black) for a halo of mass $M= 10^{12}\,h^{-1} M_{\odot}$ corresponding to $\nu=1$, and mark its virial radius $r_v=319\,h^{-1} {\rm{kpc}}$.}
\label{fig:density_nfw_kspace_eta0} 
\end{figure}
Feedback mechanims are more complex to model than adiabatic contraction. Objects like supernovae and AGN release large quantities of energy into their environments. Accounting for the former involves the cumulative effect of multiple sources heating surrounding gas, which then expands to virial radius scales \citep{Pontzen&Governato12, Lagos13}. The latter arises from gas infall from an accretion disk onto a central supermassive black hole. Radiative and kinetic modes determine whether AGN heat their environment through isotropic radiative transfer or highly directional jets respectively. In both cases a large range of scales from subparsecs up to megaparsecs are influenced, so the effect cannot be modelled analytically \citep{Schaye10, vandaalen11, Martizzi14}. The impact on halo structure is also dependent on mass. Simulations have shown \citep[e.g.,][]{Pontzen&Governato12, Teyssier13, Martizzi13} that similar mechanisms can describe the expulsion of gas from the central regions of both lower and higher mass haloes by AGN. In the former case haloes are subjected to stronger expulsions, resulting in the loss of substantial baryonic matter. Larger mass haloes are not so devastated by violent feedback, merely bloating outwards as heated gas expands through the structure. 

\citetalias{Mead15} accounts for the scale and mass dependence of feedback by transforming the scale of the halo window function according to
\begin{equation}
u\left(k\mathopen{|}\mathclose M\right)\longrightarrow u\left(\nu^{\eta}k\mathopen{|}\mathclose M\right),
\end{equation}
where
\begin{equation}
\nu\equiv \frac{\delta_c}{\sigma\left(M\right)}
\end{equation}
is the ratio of the collapse overdensity to the standard deviation of the density field, smoothed over a mass scale, $M$. As shown in Figure~\ref{fig:density_nfw_kspace_eta0}, more positive values of $\eta$ increasingly bloat higher mass haloes (characterized by $\nu> 1$) while lower mass haloes ($\nu < 1$) are left relatively reduced by gas being fully expelled. A non-zero value of $\eta$ was also required to make empirical corrections to the halo bloating to ensure accurate power spectra when fitting to dark-matter-only simulations. 

When fitting power spectra to COSMIC EMU simulations it was found (see Table 2 in \citetalias{Mead15}) that a number of parameters required redshift-dependent modifications. This includes $\eta$ which is decomposed into a constant, $\eta_0$, that controls the degree of feedback, and a fixed dependence on $\sigma_8\left(z\right)$ such that
\begin{equation}
\eta=\eta_0-0.3\,\sigma_8\left(z\right).
\end{equation}
We use $\eta_{0,\rm{fid}}=0.603$ as our fiducial value, as this is determined by \citetalias{Mead15} to best fit COSMIC EMU spectra. Fits to OWLS simulations lie within the range $0.5 \leq \eta_0 \leq 0.8$, which could be used to define a prior.

\subsubsection{Inner Cores}
\label{subsec:baryoncores}
A long-running debate about the nature of the inner most region of the halo motivates incorporating inner cores into the baryon-halo model. The cusp-core problem arises from a discrepancy between N-body simulations that predict the divergent $\rho \propto r^{-1}$ NFW cusp in halo centres \citep[see][]{Dubinski&Calberg91, NFW97}, and observations like dwarf galaxy rotation curves that indicate constant density cores of the order of a few $\rm{kpc}$ \citep{deNaray08, Walker&Pen11, Oh11b}. Explanations for this fall into three categories: 1) simulations are systematically neglecting some aspect of structure formation; 2) replacing traditional CDM with e.g., fuzzy dark matter in the form of ultra-light axions (alternatively, self-interacting dark matter or warm dark matter) can generate cores of a few $\rm{kpc}$  \citep[see e.g.,][]{Marsh15, Zhang16}; or 3) baryonic processes flatten cusps into cores \citep{ deBlok02, Pontzen&Governato12, Pen12}. A prominent proposal \citep{Pontzen&Governato12} is that supernovae feedback transfers sudden, repeated bursts of energy to surrounding gas, causing oscillations in the central gravitational potential. This in turn induces rapid orbits of dark matter particles that flatten the cusp \citep[see also][]{Read&Gilmore05, Governato12}. However, the core physics is ultimately of limited relevance here compared to understanding the impact that the phenomenology has on large-scale structure probes. 

There are various ways to introduce cores analytically within an NFW-like profile \citep[see e.g.,][]{Einasto65, Zhao96, Navarro04}. We opt for the simplest extension,
\begin{equation}
\rho\left(r\right)=\frac{\rho_N}{\left(\frac{r+r_b}{r_s}\right)\left(1+\frac{r}{r_s}
\right)^2}.
\label{eq:rhocore}
\end{equation} 
This formalism has also been employed by \citet{Pen12}, though here we denote the baryon-induced core radius as $r_b$. Setting $r_b$ to zero reduces the profile to normal NFW. To retain the advantages of a semi-analytical halo model it is useful for a modified profile to have an analytic Fourier transform. In Appendix~\ref{appendix:fouriertransforms} we show that the model possesses this property. It is also possible to introduce a halo mass dependence, for example by allowing $r_b\propto{r_s}$. However, this would entail a dependence in turn on the halo concentration and therefore a possible degeneracy with $A_B$. Instead we assume $r_b$ is determined by some combination of processes largely independent from specific halo properties. It should be emphasized that without introducing a more robust physical motivation in our modelling the accuracy of results will be limited. However, our main concern in this work is including the generic feature of an inner core to explore its impact on the halo profile and matter power spectrum. This can provide useful insight into the potential consequences of marginalizing over uncertainty in $r_b$, but should still be treated as a broad, first-pass implementation.

\begin{figure}
\includegraphics[width=\columnwidth]{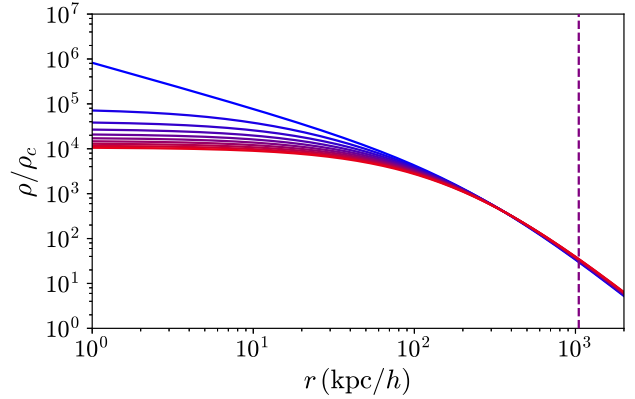}
\caption{Modified NFW profile for a halo of mass $M=3\times 10^{14}\, M_{\odot}$ featuring an inner core radius, $r_b$, in the range $r_b=0-100\,h^{-1} {\mathrm{kpc}}$ at $z=0$. Blue curves represent profiles closer to standard NFW while red curves indicate prominent baryon cores. The purple dashed line marks the virial radius, $r_v=1050\, h^{-1} {\rm{kpc}}$.}
\label{fig:density_nfw_core} 
\end{figure}

In Figure~\ref{fig:density_nfw_core} we plot a range of cores up to $r_b=100\, h^{-1}\mathrm{kpc}$ to emphasize the deviation from an NFW profile at small scales. Once the scale is reduced to $r_b$ the density turns off from the NFW branch and becomes constant. The profile is therefore increasingly suppressed by larger inner cores. 

\vspace{-6.2mm}

\section{Impact of Baryons}
\label{sec:impact}
In this section we examine the effect of varying each parameter in $\left(A_B,\eta_0,r_b,\Omega_m,\Omega_b,h,n_s,\sigma_8,w_0,w_a\right)$ on the matter power spectrum and its weak lensing projection. This informs our interpretation of a Fisher analysis of the impact that uncertainty in the baryon parameters has on cosmological parameter constraints for a Stage IV survey. Our primary focus is to understand the degradation of the dark energy FOM due to baryon degeneracies with $w_0$ and $w_a$.

\subsection{Weak Gravitational Lensing}
\label{subsec:wgl}
We take the number distribution of galaxies for a space-based, Euclid-like lensing survey to be \citep{Laurejis11},
\begin{equation}
n\left(z\right) \propto z^2\exp\left[-\left(\frac{z}{z_0}\right)^{\frac{3}{2}}\right],
\end{equation}  
\label{eq:n(z)}
\\
where $z_0=0.636$. The number of source galaxies within $\mathrm{d}\chi$ of comoving position $\chi$ is therefore $\mathrm{d}\chi n\left(\chi\right)$. The weak lensing convergence power spectrum is given by
\begin{equation}
C_{\ell,ij} = \frac{9}{4}\Omega_m^2\left(\frac{H_0}{c}\right)^4\int_0^{\chi_{\mathrm{max}}}\,\mathrm{d}\chi\, \frac{g_i\left(\chi\right)g_j\left(\chi\right)}{a^2\left(\chi\right)}P\left(k=\frac{\ell}{f_K\left(\chi\right)},\chi\right),
\label{eq:convergencepower}
\end{equation}
where $\left(i,j\right)$ denote different tomographic redshift bins, $f_K\left(\chi\right)$ is the comoving angular distance, and $g_i\left(\chi\right)$ is the total weighting function over the distribution of sources and their relative distance from lenses. This is computed up to the survey limit, $\chi_{\rm{\max}}$, according to
\begin{equation}
g_i\left(\chi\right) = \int_{\chi}^{\chi_{\mathrm{max}}}\mathrm{d}\chi^{\prime}\, n_i\left(\chi^{\prime}\right) \frac{f_K\left(\chi-\chi^{\prime}\right)}{f_K\left(\chi^{\prime}\right)}.
\end{equation}
For each bin, photometric redshift errors are accounted for by convolving the full source distribution with the probability distribution of galaxies at $z$ being measured at redshift $z_{ph}$ such that
\begin{equation}
n_i\left(z\right) = \frac{n\left(z\right)\, \int_{z_{i,-}}^{z_{i,+}}\mathrm{d}z_{ph}\, p_{ph}\left(z_{ph}|z\right)}{\int_{z_{\rm{min}}}^{z_{\rm{max}}}\mathrm{d}z^{\prime}\,n\left(z^{\prime}\right)\, \int_{z_{i,-}}^{z_{i,+}}\mathrm{d}z_{ph}\,  p_{ph}\left(z_{ph}|z^{\prime}\right)}, 
\end{equation}
in which we have normalized over the bin. A common form for the probability distribution is \citep{Ma06, Taylor07}
\begin{multline}
p_{ph}\left(z_{ph}|z\right) = \frac{1}{\sqrt{2\pi}\sigma_z\left(1+z\right)}\exp\left\{-\left[\frac{z-z_{ph}}{\sqrt{2}\sigma_z\left(1+z\right)}\right]^2\right\}, 
\end{multline}
where $\sigma_z=0.05$ is chosen for the photometric redshift error.

\begin{figure*}
\includegraphics[width=\textwidth]{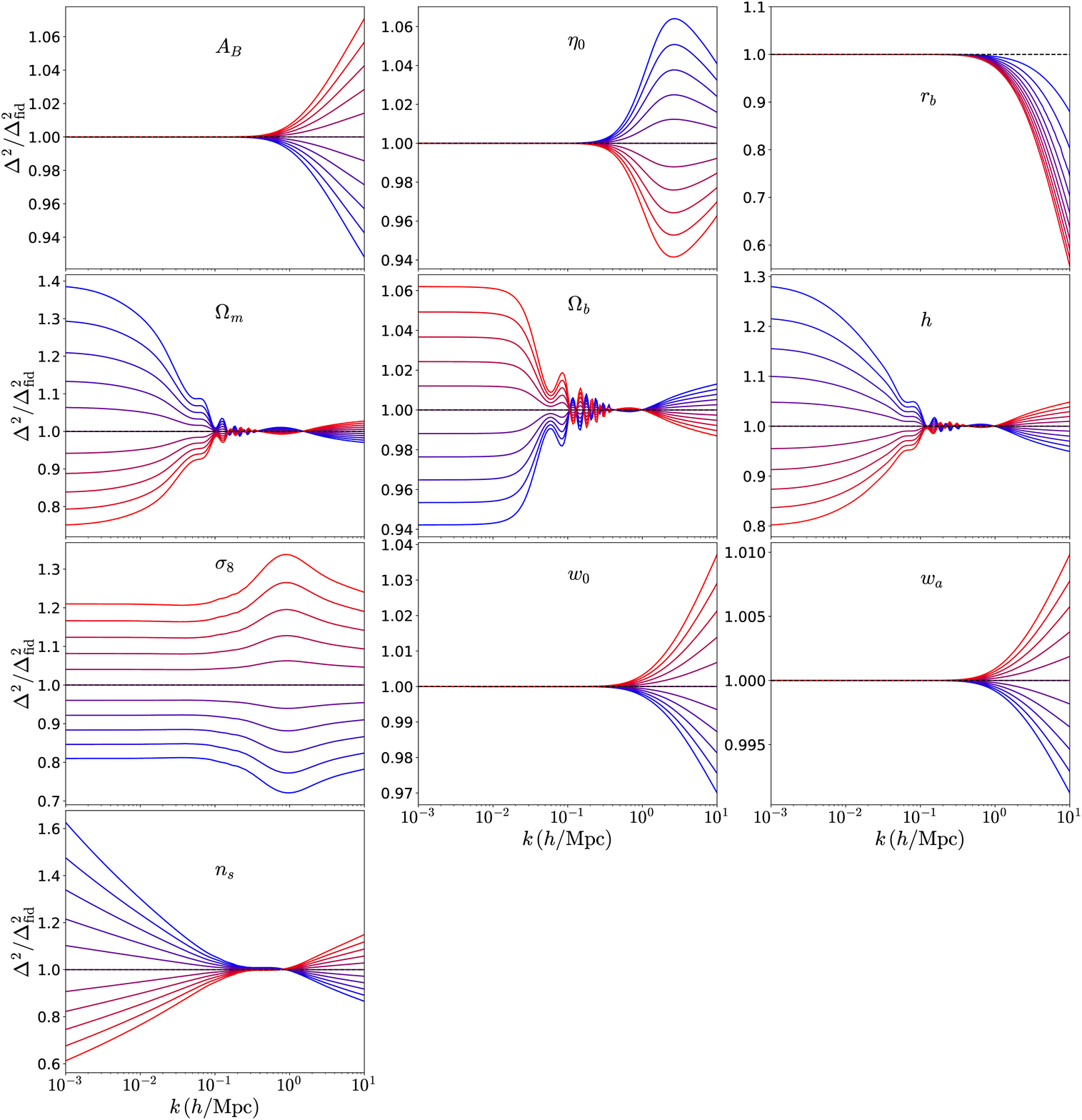}
\caption{The ratio of matter power spectra at $z=0$ for different iterations of parameters in 
$\Theta=\left(A_B,\eta_0,\Omega_m,\Omega_b,h,\sigma_8,n_s,w_0,w_a,r_b\right)$, with respect to a fiducial power spectrum computed with parameter values found by \citet{Planck15}. Bluer (redder) curves correspond to lower (higher) values for parameters in the range $0.9\,\Theta_{\rm{fid}} < \Theta < 1.1\,\Theta_{\rm{fid}}$, except in the case of the dynamic dark energy parameter which varies between $-0.1<w_a<0.1$, and $r_b$ which is varied between core sizes of $r_b=0-100\, h^{-1} \mathrm{kpc}$ and plotted with respect to the fiducial $r_b=0\, h^{-1} \mathrm{kpc}$.}
\label{fig:multipow_matter_z0}
\end{figure*}

\begin{figure*}
\includegraphics[width=\textwidth]{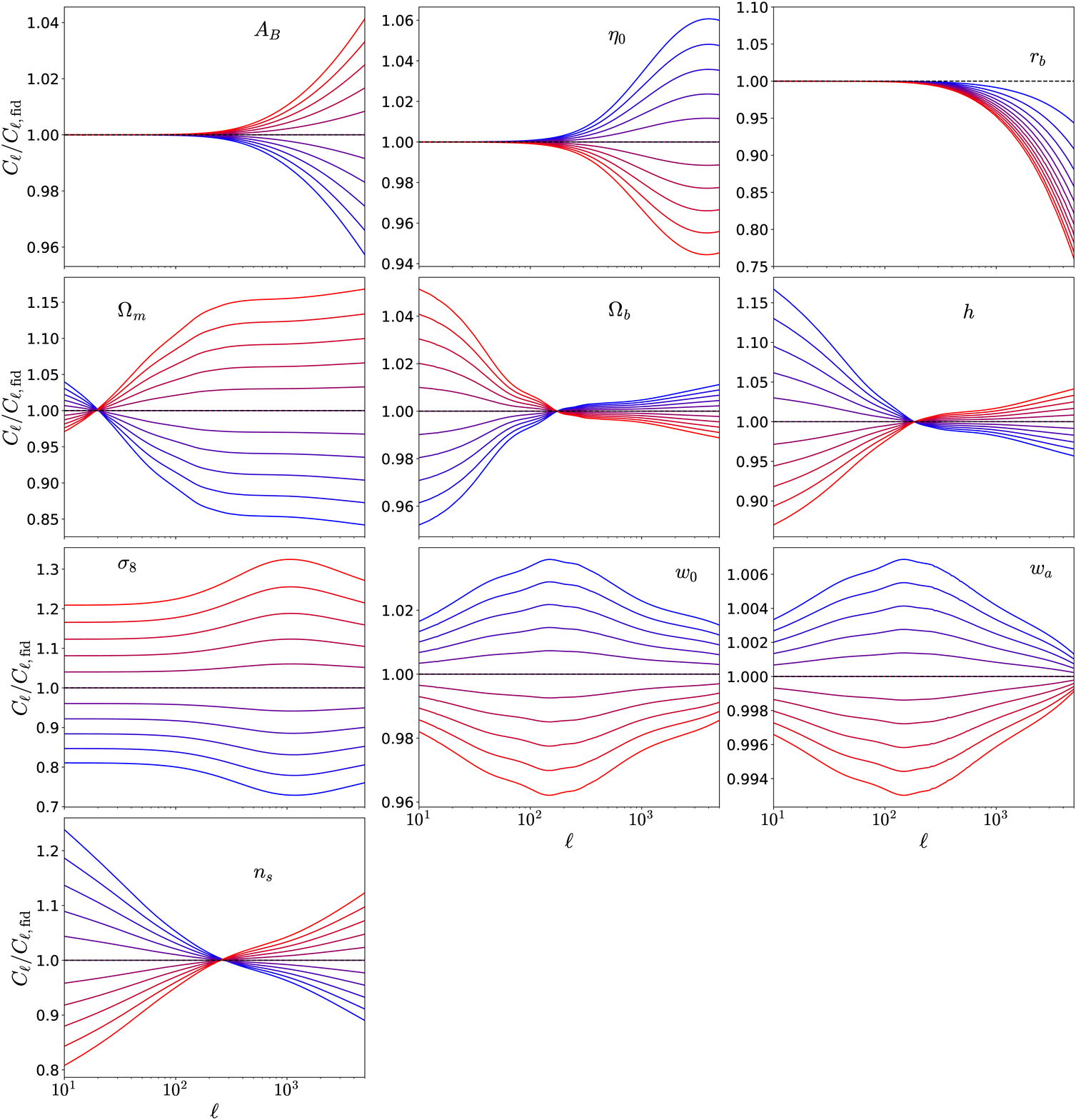}
\caption{The ratio of weak lensing convergence power spectra in a $0.9<z<1.1$ redshift bin for different parameters in $\Theta=\left(A_B,\eta_0,\Omega_m,\Omega_b,h,\sigma_8,n_s,w_0,w_a,r_b\right)$, with respect to a fiducial power spectrum computed with \citet{Planck15} parameter values. Bluer (redder) curves correspond to lower (higher) values for parameters in the range $0.9\,\Theta_{\rm{fid}} < \Theta < 1.1\,\Theta_{\rm{fid}}$, except in the case of the dynamic dark energy parameter which varies between $-0.1<w_a<0.1$, and $r_b$ which is varied between core sizes of $r_b=0-100\, h^{-1} \mathrm{kpc}$ and plotted with respect to the fiducial $r_b=0\, h^{-1} \mathrm{kpc}$.}
\label{fig:lensing_power_response}
\end{figure*}

\subsection{Parameter Accuracy Forecasts}
\subsubsection{Power spectra responses}
We can examine the sensitivity of our cosmological probe to baryons and cosmology by evaluating power spectra responses to varying each parameter with respect to its fiducial value while fixing the other parameters. Responses for $\Delta^2\left(k\right)$ at redshift $z=0$ are shown in Figure~\ref{fig:multipow_matter_z0}, and for $C_{\ell}$ in Figure~\ref{fig:lensing_power_response}. 

The most subtle response to baryons is for $\eta_0$, where averaging over the scale- and mass-dependent influences produces a peak response. The bloating impact on higher mass haloes dominates over the reduction effect on lower mass ones, so lower values of $\eta_0$ overall enhance $\Delta^2\left(k\right)$. The peak is also a function of the evolution of halo populations, occurring for smaller scales at earlier times. For $r_b$, we plot the ratio of power for cores up to $100\, h^{-1}\mathrm{kpc}$ with respect to the case of a cusp, $r_b=0 \, h^{-1} \mathrm{kpc}$. As expected, larger inner cores increasingly damp $\Delta^2\left(k\right)$ on small scales. Adiabatic contraction produces the opposite effect. Increasing $A_B$ boosts $\Delta^2\left(k\right)$ on non-linear scales, corresponding to enhanced density profiles in this regime. 

Figure~\ref{fig:multipow_matter_z0} indicates that $w_0$ and $w_a$ are degenerate with $A_B$ and $r_b$ at $z=0$. At earlier times the degeneracy is broken because the $A_B$ and $r_b$ impacts are largely redshift-independent while increasing $w_0$ and $w_a$ uniformly enhances $\Delta^2\left(k\right)$ over all scales for any $z>0$ (e.g., see Appendix~\ref{appendix:matterpowz05} for $\Delta^2\left(k\right)$ responses at $z=0.5$). This is because, for less negative values of $w$, dark energy becomes energetically relevant earlier. By $z=0$ there has been more acceleration and therefore greater suppression of structure growth, so $\Delta^2\left(k\right)$ is boosted for fixed $\sigma_8$.

For lensing, integrals of $\Delta^2\left(k\right)$ along the line of sight average over these redshift-dependent effects. Varying $w_0$ and $w_a$ now induces the opposite response for $C_{\ell}$ (see Figure~\ref{fig:lensing_power_response}) than for $\Delta^2\left(k\right)$. Notably there is also a broad peak on large scales ($\ell \sim 100$). This is because dark energy influences cosmological distances and therefore rescales the lensing weight functions. More negative $w$ increases this geometric contribution to the lensing signal \citep{Huterer01}, boosting $C_{\ell}$ on all scales. On non-linear scales this is damped by the opposite influence from the growth of structure, which enters through $\Delta^2\left(k\right)$. For the smallest $\ell$, linear $k$ can only be accessed at larger distances and therefore earlier times, when more positive values of $w$ boost $\Delta^2\left(k\right)$. This also has a mitigating effect on the influence of geometry, leading to the broad response peak. Our results are consistent with those of \citet{Matilla17}, who thoroughly examine the competing effects of geometry and growth on the sensitivity of lensing observables to $\Omega_m$ and $w$.      

\vspace{-2.2mm}

\subsubsection{Fisher formalism}
\label{subsubsec:fisher}
\vspace{-2.2mm}
We can estimate the covariance of parameter values, $\Theta=\left(\theta_1,...,\theta_N\right)$, from a data set, $\bmath{x}$, by using the Fisher formalism. For a Gaussian posterior, the inverse of the parameter covariance matrix is given by the expectation value of the curvature of the likelihood function, $L\left(\bmath{x}\lvert\Theta\right)$, at the fiducial, most likely parameter values, $\Theta_{\rm{fid}}$ \citep*[e.g.,][]{Tegmark97}. This quantity is the Fisher information matrix,
\vspace{-1.48mm}
\begin{equation}
F_{\alpha\beta} \equiv \Big\langle \frac{\partial^2 \mathcal{L}}{\partial\theta_{\alpha}\partial\theta_{\beta}} \Big\rangle,
\end{equation} 
where $\mathcal{L}=-\ln L$. Under the Gaussian approximation,
\begin{equation}
L=\frac{1}{\sqrt{\left(2\pi\right)^N\det \bmath{C}}}\exp\left[-\frac{1}{2}\left(\bmath{x}-\boldsymbol{\mu}\right)\bmath{C}^{-1}\left(\bmath{x}-\boldsymbol{\mu}\right)^{T}\right],
\end{equation}
where $\bmath{C}$ is the data covariance matrix and $\boldsymbol{\mu}=\langle\bmath{x}\rangle$ is the mean data, the Fisher matrix can be written
\begin{equation}
F_{\alpha\beta} = \frac{1}{2} \mathrm{Tr}\left[\bmath{C}^{-1}\bmath{C}_{,\alpha}\bmath{C}^{-1}\bmath{C}_{,\beta} + \bmath{C}^{-1}M_{\alpha\beta}\right].
\end{equation}
The final term, $M_{\alpha\beta}=\boldsymbol\mu_{,\alpha}\boldsymbol\mu_{,\beta}^T+\boldsymbol\mu_{,\beta}\boldsymbol\mu_{,\alpha}^T$, is the expectation value of the second derivative of the data matrix $\left(\bmath{x}-\boldsymbol{\mu}\right)\left(\bmath{x}-\boldsymbol{\mu}\right)^{T}$ under Gaussian conditions. Derivatives with respect to parameters are denoted by $,\alpha \equiv \partial/\partial\Theta_{\alpha}$. 

Our observable is the weak lensing convergence power spectrum, which is approximated to be Gaussian. The corresponding Fisher matrix is \citep[][]{Tegmark97, Takada&Jain04}
\begin{equation}
F_{\alpha\beta} = \frac{1}{2}f_{\rm{sky}}\sum_{\ell}  \left(2\ell + 1\right) \sum_{\left(ij\right)}\sum_{\left(pq\right)} C^{ij}_{\ell,\alpha}C^{pq}_{\ell,\beta} \left[\rm{Cov}^{-1}\right]_{\ell,\left(ij\right),\left(pq\right)},
\label{eq:fishertrace}
\end{equation}
in which the spherical harmonic $\ell$ and $m$ modes are summed over, and $f_{\rm{sky}}$ is the fraction of sky accessible to the survey. The auto- and cross-correlations of observed power in redshift bins $i,j,p,q=\left(1,...,N_{\rm{bin}}\right)$ are captured by the covariance matrix,
\begin{equation}
{\rm{Cov}}_{\ell,\left(ij\right),\left(pq\right)} = \hat{C}^{ip}_{\ell}\hat{C}_{\ell}^{jq} + \hat{C}_{\ell}^{iq}\hat{C}_{\ell}^{jp}.
\end{equation}
Contributions from different modes are treated as separable so that the matrix is block diagonal in $\ell$. The full observed power spectrum is constructed by adding the shape noise, $\sigma_e=0.3$, when averaged over the number of galaxies to the auto-correlations of power within each bin. Provided there is no intrinsic alignment to account for, we write
\begin{equation}
\hat{C}_{\ell,ij} = {C}_{\ell,ij} + \frac{\sigma_e^2}{n_i}\delta_{ij},
\end{equation}
where $n_i$ is the number density of galaxies in redshift bin $i$.  

A 2-parameter confidence region is finally determined by inverting the Fisher matrix to marginalize over the other parameters, and extracting the resulting 1-$\sigma$ errors, $\sigma_{\alpha\beta}^2=\left[F^{-1}\right]_{\alpha\beta}$. 

\subsection{Results}
We construct $\left(A_B,\eta_0,r_b,\Omega_m,\Omega_b,h,n_s,\sigma_8,w_0,w_a\right)$ Fisher matrices for a Euclid-like survey. In Table~\ref{table:cosmicemurange}, we state the survey parameters specified by the Euclid survey report \citep{Laurejis11}. $N_{\mathrm{bin}}=10$ redshift bins are chosen in the range $0<z<2$ such that each bin contains an equal number density of galaxies,
\begin{equation}
\mathcal{N}=\frac{1}{N_{\mathrm{bin}}}\int_0^{z_{\rm{max}}}\mathrm{d}z\, n\left(z\right),
\end{equation}
where $n\left(z\right)$ is the redshift distribution of the number density. A large range of scales from $\ell_{\mathrm{min}}=10$ to $\ell_{\mathrm{max}}=5000$ are covered so in practice we compute the summation in equation~\eqref{eq:fishertrace} at logarithmic intervals.
\begin{table}
\centering
\begin{tabular}{|c|c|}
\hline
Parameter & Euclid value \\
  \hline 
  $A_{\rm{sky}}$ & $15,000\, \mathrm{deg}^2$ \\
 
  $z_{\rm{min}}$ & $0.$ \\
 
  $z_{\rm{max}}$ & $2.0$ \\
 
$z_{\rm{med}}$ & $0.9$ \\
$N_{\rm{bin}}$ & $10$ \\
$n_{\rm{gal}}$ & $30$ gal/arcmin$^2$ \\
$\sigma_z$ & $0.05$ \\
$\sigma_e$ & $0.3$ \\ 
$l_{\rm {min}}$ & $10$ \\  
$l_{\rm {max}}$ & $5000$ \\

  \hline
\end{tabular}
\caption{Survey parameters for a Euclid-like space mission, including the area $A_{\rm{sky}}$ of sky probed, the redshift range and median redshift value $z_{\rm{med}}$, the number of redshift bins $N_{\rm{bin}}$, the number density of galaxies surveyed, $n_{\rm{gal}}$, the photometric redshift error $\sigma_z$, the intrinsic ellipticity $\sigma_e$, and the range of accessible harmonic wavenumbers.}
\label{table:cosmicemurange}
\end{table}
\begin{figure}
\includegraphics[width=\columnwidth]{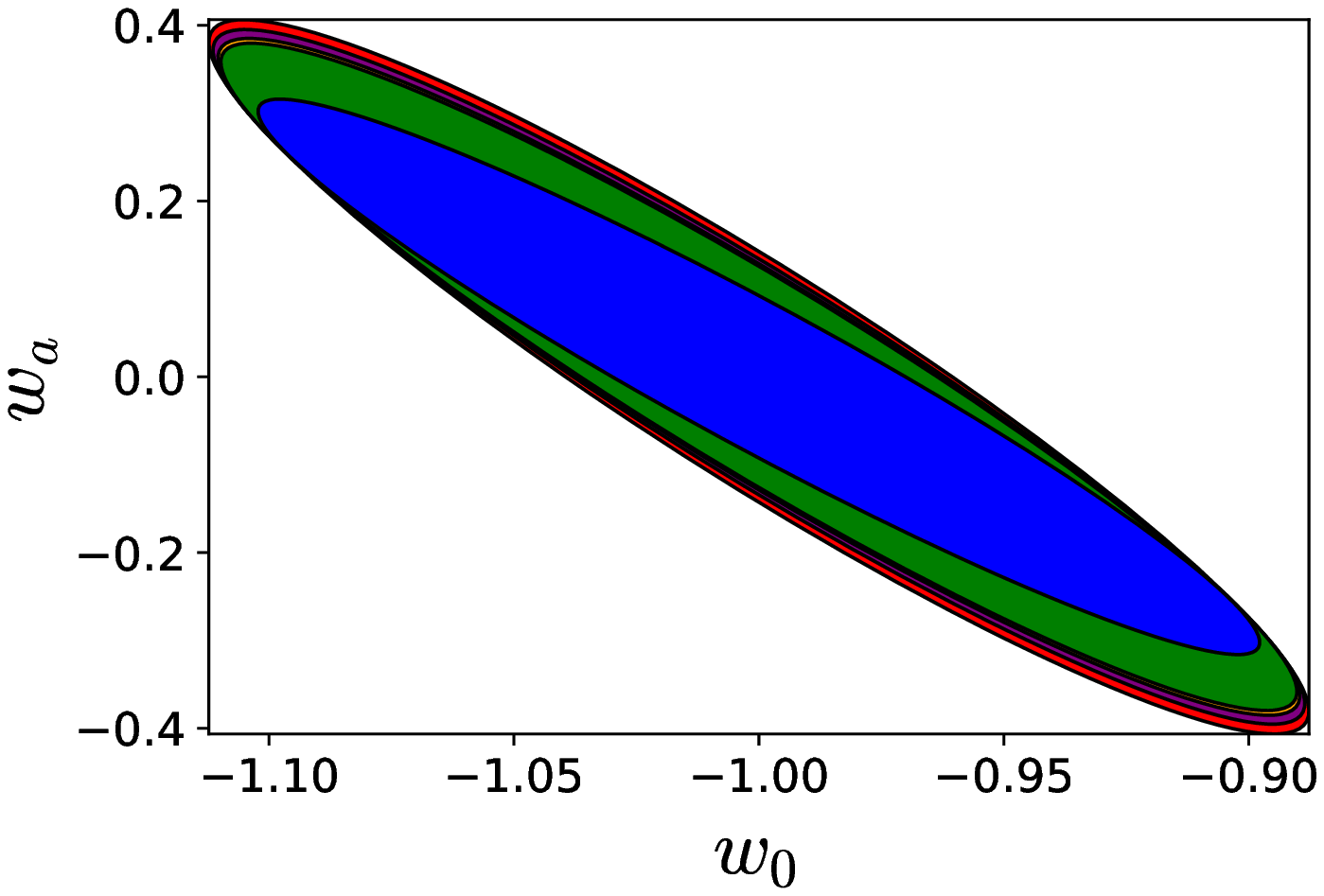}
\caption{1-$\sigma$ 2-parameter confidence ellipses for $w_0$ and $w_a$. In each case, $\Omega_m,\Omega_b,h,n_s,\sigma_8$ have been marginalized over. We show results when all baryon parameters are fixed to their fiducial values (blue); one baryon parameter fixed to their fiducial value, $A_B$ (orange), $\eta_0$ (green), $r_b$ (purple); and all baryon parameters marginalized over (red; just visible as the largest ellipse).}
\label{fig:lens_w0_wa_istnorm_zbin} 
\end{figure}
\begin{figure}
\includegraphics[width=\columnwidth]{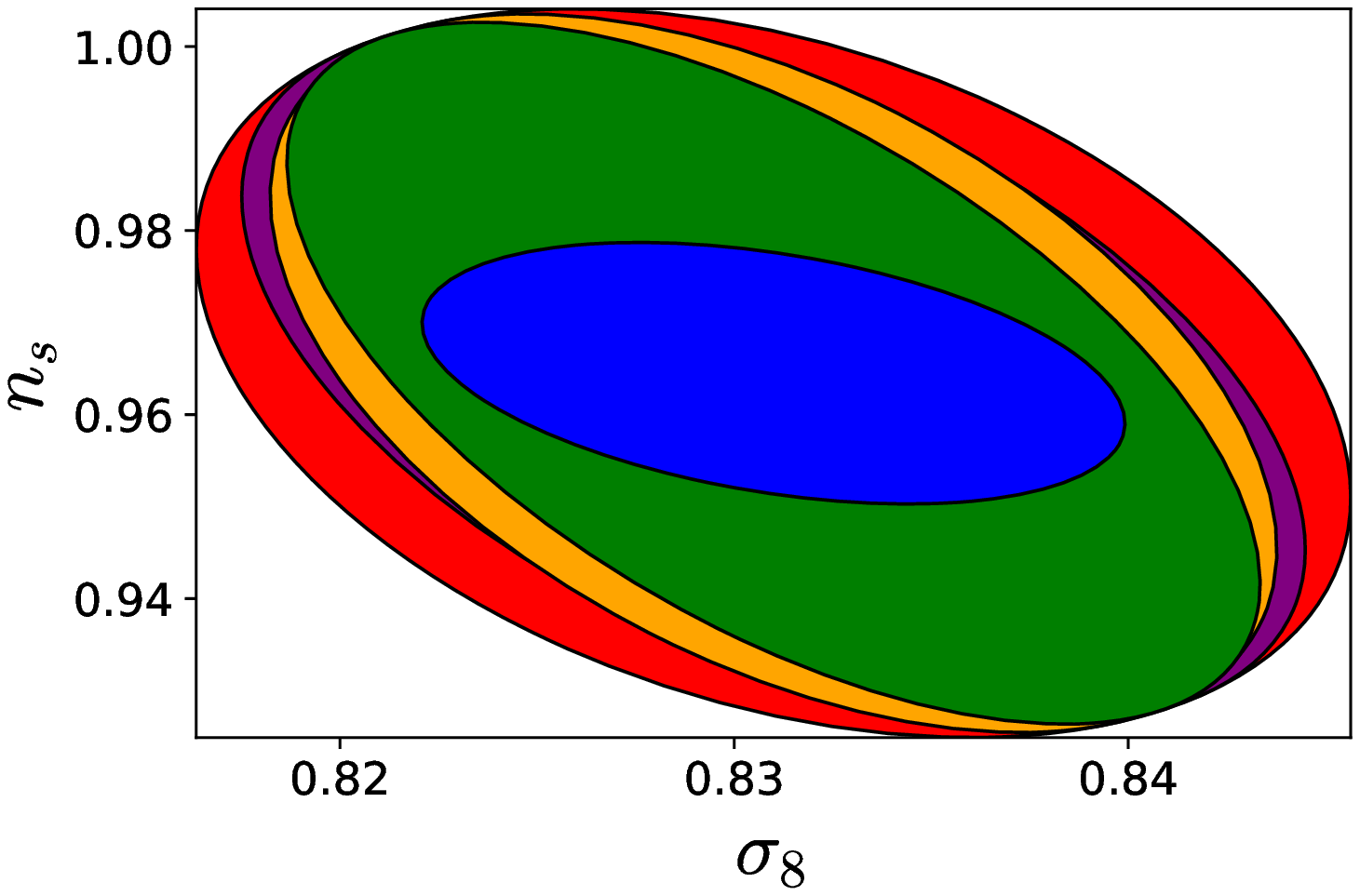}
\caption{1-$\sigma$ 2-parameter confidence ellipses for $\sigma_8$ and $n_s$. In each case, $\Omega_m,\Omega_b,h,w_0,w_a$ have been marginalized over. We show results when all baryon parameters are fixed to their fiducial values (blue); one baryon parameter fixed to their fiducial value, $A_B$ (orange), $\eta_0$ (green), $r_b$ (purple)); and all baryon parameters marginalized over (red).}
\label{fig:lens_sig8_ns_omm_h_planck_eqzbin} 
\end{figure}

\begin{table}
\centering
\begin{tabular}{c|c|c|c}
  \hline
& 1-$\sigma$  & 1-$\sigma$  & $f_{\rm{deg}}$ \\
& (no baryon marg.) & (inc. baryon marg.) &  \\
\hline
$w_0$ & 0.0673 & 0.0739 & 1.10 \\
$w_a$ & 0.208 & 0.267 & 1.29 \\
\hline
$\sigma_8$ & 0.00587 & 0.00964 & 1.64 \\
$n_s$ & 0.00935 & 0.0260 & 2.79 \\
\hline  
\end{tabular}
\caption{1-$\sigma$ error forecasts for dark energy and selected cosmological parameters for a Euclid-like survey, without and including marginalization over baryon parameters. The ratio, $f_{\rm{deg}}$, of errors with baryons marginalized over to those without baryon marginalization quantifies the degradation.}
\label{table:barerrors}
\end{table}
\begin{table}
\centering
\begin{tabular}{c|c|c|c}
  \hline
  & $\rm{FOM}$  & $\rm{FOM}$ & $R_{\rm{FOM}}$ \\
& (no baryon marg.) & (inc. baryon marg.) &   \\
\hline
$w_0$-$w_a$ & 106 & 62.4 & 1.70     \\
$n_s$-$\sigma_8$ & 8540 & 1830 & 4.65  \\
\hline
\end{tabular}
\caption{Figures of merit for $w_0$-$w_a$ and $n_s$-$\sigma_8$ without and including marginalization over baryonic physics. We include the reduction factor, $R_{\rm{FOM}}$, calculated as the ratio of the FOM without baryon marginalization to the FOM with baryon parameters kept fixed.}
\label{table:basicfom}
\end{table}
The full set of 2-parameter 1-$\sigma$ confidence ellipses from this analysis is shown in Appendix~\ref{appendix:confellip}. We obtain results where different combinations of baryon parameters are systematically fixed, or marginalized over alongside the cosmological parameters. In Figure~\ref{fig:lens_w0_wa_istnorm_zbin} we show the dark energy error forecasts. There is a baryon degradation factor (computed as the ratio of the 1-$\sigma$ errors when marginalizing over baryons to those when baryon parameters are fixed), $f_{\rm{deg}}= 1.10$, on the $w_0$ error, and a more substantial degradation of $f_{\rm{deg}}=1.29$ on the $w_a$ error (see Table~\ref{table:barerrors}). These compound such that the dark energy FOM is reduced by a factor $R_{\rm{FOM}}=1.70$.

When baryons are fixed to their fiducial values, the $w_0$-$w_a$ parameter space can be constrained because the $w_0$-$w_a$ degeneracy is broken by two sources. The first is differences in the scale dependence of the lensing power response to varying $w_0$ and $w_a$. In  Figure~\ref{fig:lensing_power_response} we illustrate that on non-linear scales, for a given redshift slice, competing influences from geometry and the matter power spectrum damp the net sensitivity of the lensing power to $w_a$. By contrast, the effect of varying $w_0$ on geometry remains dominant over its impact on the matter power deeper into the non-linear regime. The second source is contributions from multiple photometric redshift bins. The evolution of the lensing power spectrum depends on the growth rate and integrals over the line of sight, which respond differently to $w_0$ and $w_a$. However, degeneracies between baryons and dark energy on non-linear scales obscure these distinctions in the Fisher analysis when baryons are marginalized over. Breaking the $w_0$-$w_a$ degeneracy now depends on evolution alone and consequently the FOM experiences significant $\sim 40$\% degradation.

In Figure~\ref{fig:lens_baryon_constraints_istnorm} we illustrate that other cosmological parameters are also vulnerable to baryon degradation, using $n_s$-$\sigma_8$ as an example. $P\left(k\right)$ and $C_{\ell}$ experience non-linear peak responses to $\sigma_8$ and $\eta_0$ (see Figures~\ref{fig:multipow_matter_z0} and~\ref{fig:lensing_power_response}). The spectral index amplifies power with scale $k>1\, h\, \mathrm{Mpc}^{-1}$ similarly to varying $A_B$ and $r_b$. These combined degeneracies reduce the $n_s$-$\sigma_8$ FOM to $\sim 20$\% of its value before baryons are marginalized over. Though this work is focused mainly on dark energy, we include this result to highlight the importance of understanding baryonic effects for constraining all cosmological parameters. We do not address bias in the measurements of $n_s$, $\sigma_8$ and other cosmological parameters, instead referring the reader to \citet{Semboloni11} who find considerable biases of up to $\sim 10$\% in their parameter estimates if the influence of baryons is neglected. This is separate to the degradation of error forecasts we find from marginalizing over baryonic effects once they have been included.

Fixing any single baryon parameter does not significantly reduce the degradation on dark energy. Setting the inner core radius to zero, for example, will yield very limited improvement. However, Figure~\ref{fig:lens_sig8_ns_omm_h_planck_eqzbin} exhibits more varied impacts on $n_s$-$\sigma_8$ constraints when selectively fixing different baryon parameters. This signals the importance of accounting for multiple baryon influences. Figure~\ref{fig:lens_baryon_constraints_istnorm} shows there are correlations between baryon parameters in Stage IV forecasts but they are not sufficiently degenerate to motivate reduction to a single parameter, as in \citet{Hildebrandt17}.

The dependence of forecasts on the choice of baryon fiducial values should also be briefly noted. For example, selecting $A_B$ and $\eta_0$ values that best fit the OWLS AGN simulation ($A_B=2.32$, $\eta_0=0.76$) results in a $\sim$10\% change in $w_0$ constraints, giving $\sigma_{w_0,\rm{AGN}}=0.104$. 

\subsubsection{Constraining Baryon Parameters}
\begin{figure}
\includegraphics[width=\columnwidth]{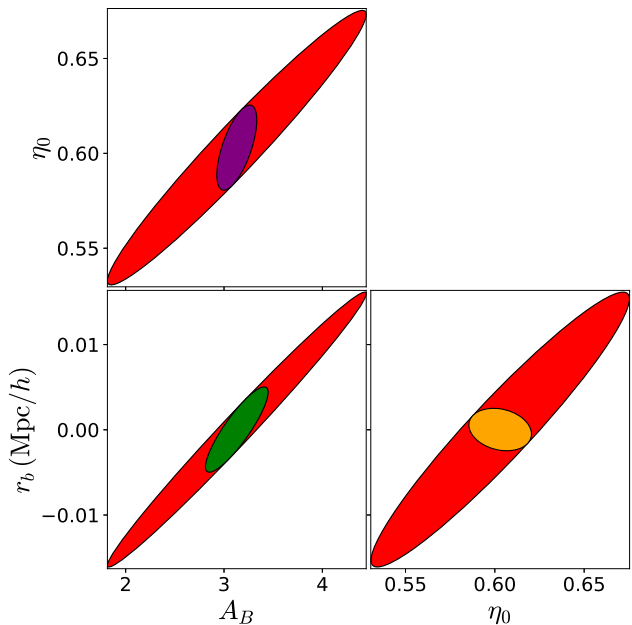}
\caption{1-$\sigma$ 2-parameter confidence ellipses for $\left(A_B,\eta_0,r_b\right)$: marginalized over all remaining parameters in $\Theta=\left(A_B,\eta_0,r_b,\Omega_m,\Omega_b,\right.$ $\left. h,\sigma_8,n_s,w_0,w_a\right)$ (red); $A_B$ fixed to its fiducial value (orange); $\eta_0$ fixed (green); and $r_b$ fixed (purple).}
\label{fig:lens_baryon_constraints_istnorm} 
\end{figure}
Our framework also allows us to quickly compute confidence limits for prospective measurements of the different baryon parameters in HMCODE by future surveys. Such information could then be useful for modelling baryons in future simulations. This is a further advantage of using multiple baryon parameters. A thorough study, also centered on the \citetalias{Mead15} parameters, is presented by \citet{MacCrann17} with respect to the Dark Energy Survey \citep{DES05, Abbott16}. Our forecasts are for Stage IV Euclid-like surveys and also include inner halo cores. In principle higher order statistics (e.g., combining two- and three-point shear statistics) could also be used to improve constraints on baryon parameters \citep{Semboloni13}. In \S~\ref{sec:modbias} we evaluate how strong any such potential priors would have to be to lead to significant improvements in the dark energy forecasts. While the constraints presented here are limited to the effects of baryons on the matter distribution, it should be noted that more direct constraints on baryonic phenomenology can be found through studying baryon fraction scaling relations with halo mass \citep{Martizzi14, Wu18}. 

In Figure~\ref{fig:lens_baryon_constraints_istnorm} we find that $A_B$ and $\eta_0$ could be constrained by a Euclid-like survey at the $50$\% and $10$\% level respectively, with 1-$\sigma$ errors $\sigma_{A_B}=0.866$ and $\sigma_{\eta_0}=0.0476$. A significant improvement is made, particularly for constraining adiabatic contraction, by fixing the inner core radius to zero, as this breaks the degeneracy between $A_B$ and $r_b$. This reduces the errors by several factors to $\sigma_{A_B,\rm{cusp}}=0.134$ and $\sigma_{\eta_0,\rm{cusp}}=0.0148$ for a cuspy halo model, while the  $A_B$-$\eta_0$ FOM is greater by a factor of $6.7$. The $A_B$-$r_b$ and $\eta_0$-$r_b$ constraints experience similar improvements when fixing the third baryon parameter in each case, further highlighting the degeneracies that occur between the different effects. It should also be noted that $\eta_0$ and $r_b$ exhibit a positive correlation despite having opposite effects on $C_{\ell}$. This is because there is also marginalization over $A_B$, which has a dominant influence compared to $r_b$. 

Surprisingly, our results imply that a Euclid-like survey could be highly sensitive to cores on the smallest scales, within $0.02 \,h^{-1}\mathrm{Mpc}$. Various axionic mechanisms like fuzzy dark matter or solitonic field configurations of self-gravitating bosons that generate halo cores are posited to exist on $\mathrm{kpc}$ scales \citep[e.g.,][]{Marsh15}. Such a preference in favour of the axion sector over baryon-induced mechanisms of core formation would be significant for the cusp-core debate. However, we do not realistically expect our model to be robust at these scales. HMCODE does not account directly for physical baryonic processes and our extension to an inner core similarly does not incorporate the behaviour of stars and gas that would dominate on these scales. The inner core is a generic, phenomenological parameterization without a robust physical motivation for how it is implemented. Therefore, constraints on $r_b$ should not be taken as a definitive forecast for the size or implied nature of an inner core. Rather, we report the finding primarily as an indication that a more sophisticated approach to incorporating inner cores could draw substantial benefits from Stage IV forecasts. 

As the scope of this work is primarily focused on dark energy constraints, the main result of this section is that only 59\% of the dark energy FOM is retained when baryons are marginalized over (see Table~\ref{table:basicfom}). It is now necessary to explore possible mitigation strategies and identify with greater precision the scales at which constraining information becomes compromised.

\section{Mitigation}
\label{sec:mitigation}
\begin{table}
\centering
\begin{tabular}{c|c|c|c}
  \hline
& 1-$\sigma$  & 1-$\sigma$  & $f_{\rm{deg}}$ \\
& (no baryon marg.) & (inc. baryon marg.) &  \\
\hline
$w_0$ & 0.0656 & 0.0725 & 1.10 \\
$w_a$ & 0.197 & 0.258 & 1.31 \\
\hline
$\sigma_8$ & 0.00531 & 0.00903 & 1.70 \\
$n_s$ & 0.00740 & 0.0247 & 3.34 \\
\hline  
\end{tabular}
\caption{1-$\sigma$ error forecasts for dark energy and selected cosmological parameters for a survey probing up to $\ell_{\rm{max}}=10000$ , without and including marginalization over baryon parameters. The final column shows the degradation factor induced in the errors by baryonic physics.}
\label{table:10000errors}
\end{table}

\begin{table}
\centering
\begin{tabular}{c|c|c|c}
  \hline
$\ell_{\rm{max}}$& $\rm{FOM}_{w_0\rm{-}w_a}$  & $\rm{FOM}_{w_0\rm{-}w_a}$  & $R_{\rm{FOM}}$ \\
& (no baryon marg.) & (inc. baryon marg.) &  \\
\hline
$5000$ & 106 & 62.4 & 1.70 \\
$10000$ & 121 & 66.5 & 1.82 \\
\hline
$r_{10000/5000}$ & 1.15 & 1.07 & 1.07 \\
\hline  
\end{tabular}
\caption{Figures of merit for $w_0\rm{-}w_a$ without and including marginalization over baryon parameters, at $\ell_{\rm{max}}=5000$ to $\ell_{\rm{max}}=10000$. The final column and row respectively show the reduction fraction, $R_{\rm{FOM}}$, of the FOM when including marginalization over baryonic physics and the ratio, $r_{10000/5000}$, of quantities computed using $\ell_{\rm{max}}=10000$ to those computed using $\ell_{\rm{max}}=5000$.}
\label{table:fomerrors}
\end{table}

\begin{figure}
\includegraphics[width=\columnwidth]{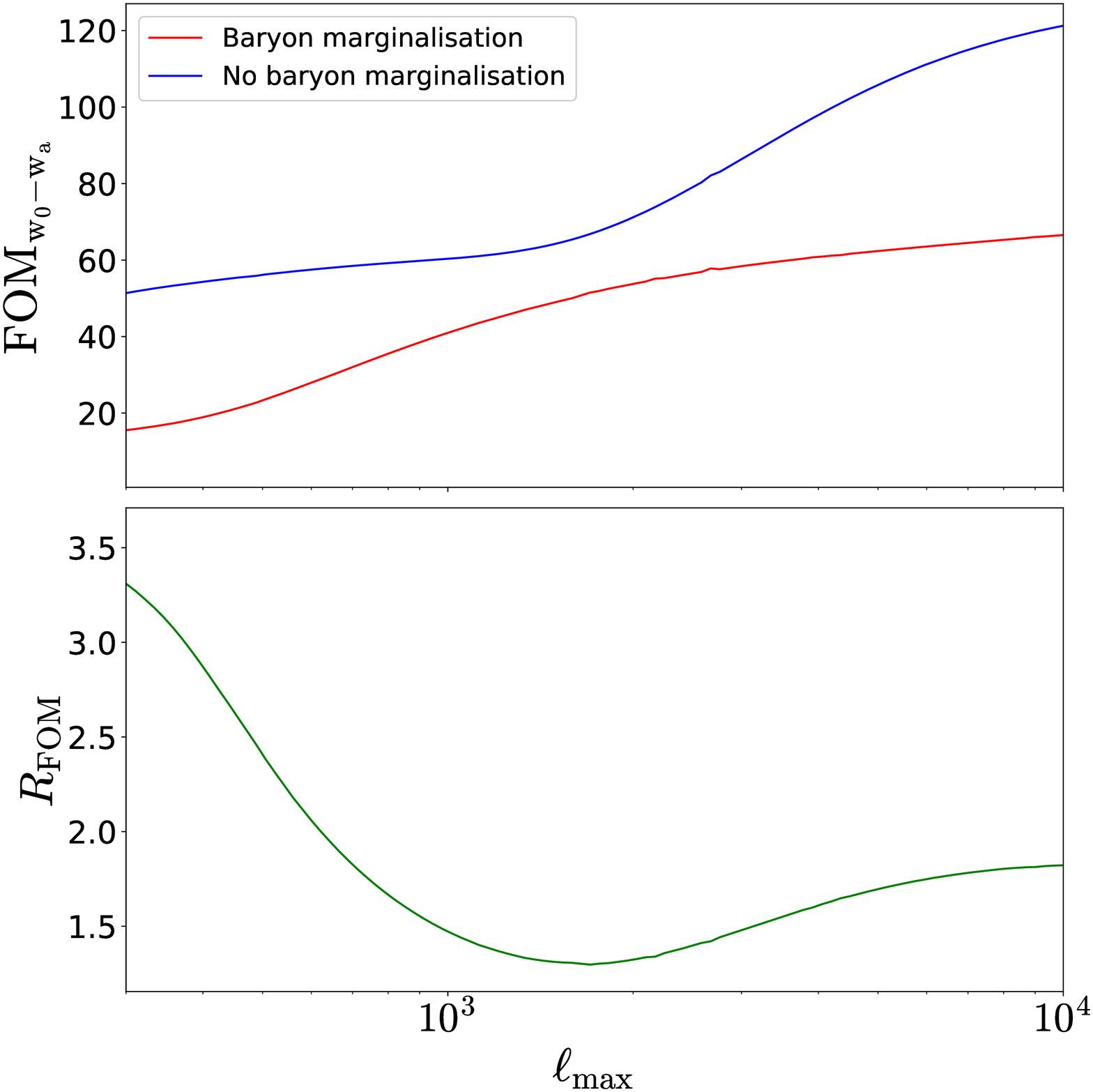}
\caption{Top panel: FOM for $w_0\rm{-}w_a$ at different $\ell$-mode cutoffs, with baryon parameters fixed to their fiducial values (blue) and marginalized over (red). Bottom panel: the reduction factor, $R_{\rm{FOM}}$ on the FOM due to baryon uncertainty.}
\label{fig:sensitivity_degradation} 
\end{figure}

\begin{figure*}
\includegraphics[width=\textwidth]{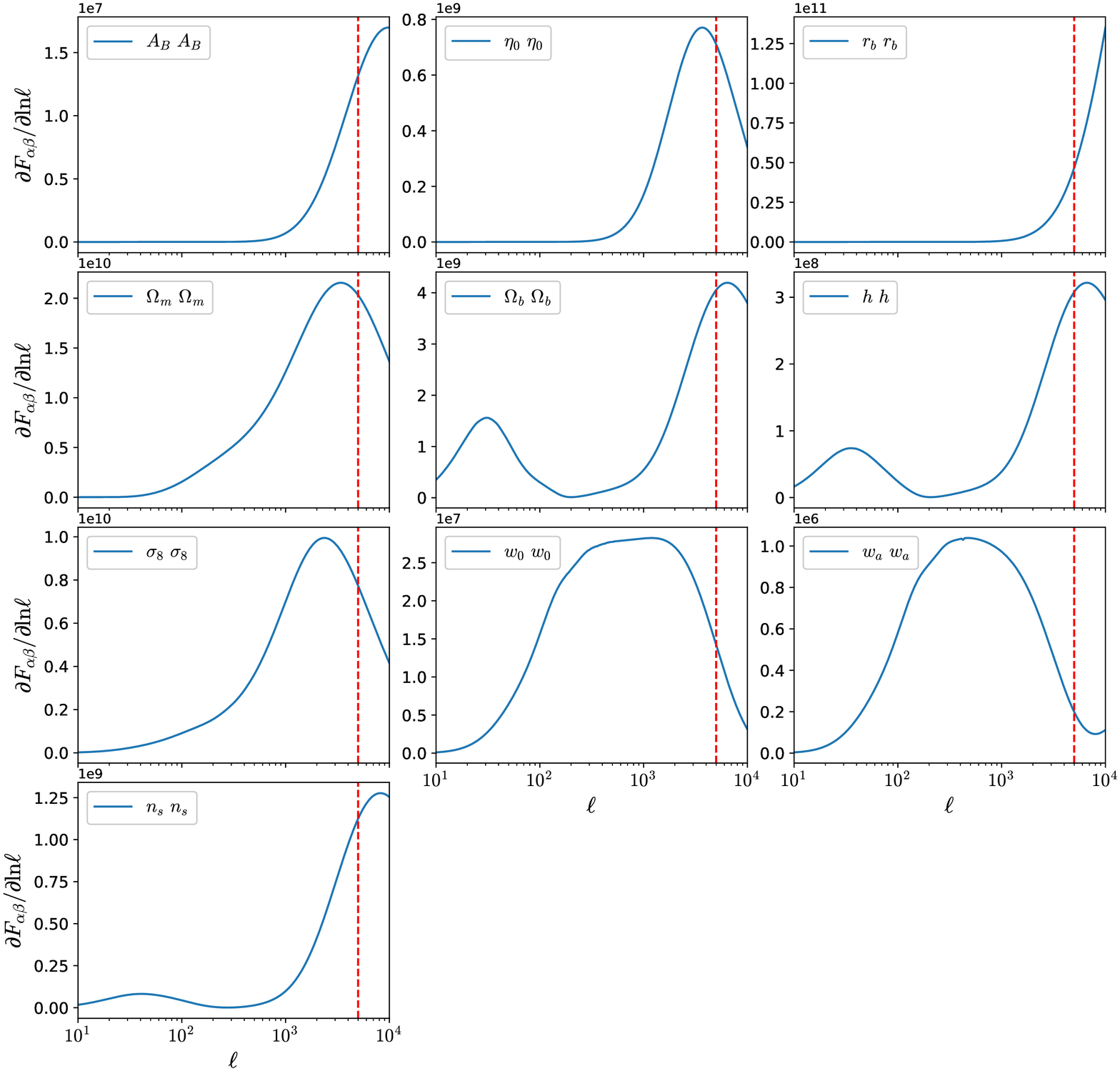}
\caption{Diagonal contributions to the Fisher information at each $\ell$-mode for each cosmological and baryon parameter. The vertical line (red, dashed) in each panel marks the conventional cutoff for Euclid-like surveys at $\ell_{\rm{max}}=5000$. The derivative with respect to $\ln \ell$ is shown to more accurately convey the integration area covered by the logarithmically spaced $\ell$-range.}
\label{fig:sensitivity_w0_allbar} 
\end{figure*}

\subsection{Fisher Information Sensitivity}
An analysis that extends further into non-linear scales might be expected to improve upon the baryon impact. Instead we see in Tables~\ref{table:10000errors} and~\ref{table:fomerrors} that doubling the survey limit from $\ell_{\rm{max}}=5000$ to $\ell_{\rm{max}}=10000$ offers a relatively minor improvement of $\sim 7$\% on the dark energy FOM. More than twice this gain is available when baryons are kept fixed. This reflects the fact that baryon degradation actually worsens at higher $\ell_{\rm{max}}$, as shown in Figure~\ref{fig:sensitivity_degradation}.

We find degradation is minimized at $\sim 20$\% for the cutoff $\ell_{\rm{max}}^{\star} \approx 1700$. The corresponding marginalized errors on $w_0$ are $\sigma_{w_0,b}^{\star} = 0.080$ and $\sigma_{w_0,nb}^{\star} = 0.074$ for marginalizing over and fixing baryons respectively. Beyond this point, the degradation worsens while the overall improvement in the FOM tails off. Therefore, it should not be assumed that simply increasing $\ell_{\rm{max}}$ is necessarily the best mitigation strategy.

The minimal baryon degradation at $\ell_{\rm{max}}^{\star}$ can be understood by considering the scale dependence of Fisher information contributions from different parameters. Figure~\ref{fig:sensitivity_w0_allbar} shows this for the diagonal terms of the Fisher matrix. The peak contributions for the baryon parameters occur almost entirely over non-linear scales. However, the dark energy contribution is evenly distributed across a range of linear and quasi-nonlinear scales of a few hundred $\ell$, quickly dropping off beyond $\ell \sim 1000$. This is due to the combination of decreasing sensitivity of lensing power to $w_0$ and $w_a$ on these scales (see Figure~\ref{fig:lensing_power_response}) and the increasing influence of shape noise. 

In Figure~\ref{fig:sensitivity_degradation} the FOM branch without baryon marginalization experiences an upturn at a greater $\ell_{\rm{max}}$ than the scale of the dark energy Fisher information peak. This is because the differences in lensing power sensitivity to $w_0$ and $w_a$ on non-linear scales help to break the $w_0$-$w_a$ degeneracy. Baryons impair this capacity, so the baryon marginalization branch relies almost entirely on responses varying with evolution to break the degeneracy. Therefore, there is no upturn, resulting in the relative degradation increasing.  

For low $\ell_{\rm{max}}$, though most of the dark energy information is available, baryons provide almost no information. The Fisher matrix is therefore close to singular at low $\ell_{\rm{max}}$ and the forecast $\mathrm{FOM}$ after marginalizing over baryons is likely inaccurate, although we expect the decrease in $\mathrm{FOM}$ with decreasing $\ell_{\rm{max}}$ to be qualitatively correct. In practice for very low values of $\ell_{\rm{max}}$ we would fix the baryon parameters, allowing them to vary only when $\ell_{\rm{max}}$ is high enough that the data is sufficiently informative that fixing them does not induce significant biases.

There is a caveat to our analysis. While calculating Fisher information using power spectra covariance matrices that are block diagonal in $\ell$ is a reasonable approximation, it is not strictly accurate. Particularly at low redshifts, non-linear modes couple and further correlate power. The traditional Fisher framework \citep{Tegmark97} can be extended to account for this additional information \citep{Kiessling11}. It is beyond the scope of this work to include the necessary non-Gaussian corrections, but we highlight that they should be considered for future extensions.

\vspace{-2.2mm}

\subsection{External Baryon Priors}
\begin{figure}
\includegraphics[width=\columnwidth]{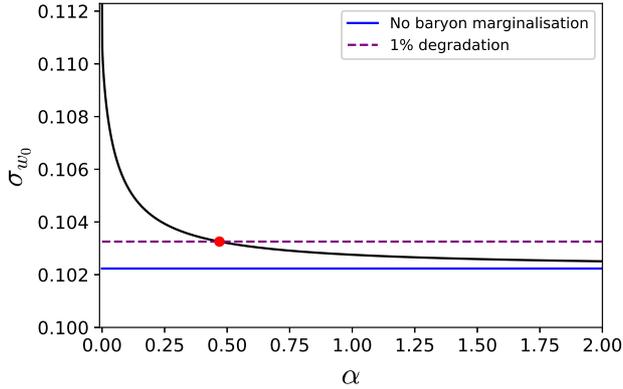}
\caption{1-$\sigma$ marginalized errors for $w_0$ as a function of the enhanced baryon Fisher information parameter $\alpha$. The blue line indicates the absolute best case scenario of no baryon uncertainty, while the purple dashed line represents an acceptable degradation threshold of $1\%$. The red dot marks the $\alpha\approx 0.47$ value required to achieve this improvement.}
\label{fig:fisher_improvement} 
\end{figure}

As little mitigation is offered by extending the minimum scale of an analysis, we can instead consider adding information from independent sources. We first explore the possibility of an external baryon prior to break degeneracies with dark energy and increase the total Fisher information available. For example, \citet{Hildebrandt17} adopt a top-hat prior, $2<A_B<4$, based on the range of fits to different OWLS simulations \citetalias{Mead15} find for $A_B$ and $\eta_0$. A stronger prior will have to come from future observations or simulations, so it is important to quantify the level of further information required. We provide a recommendation for limiting degradation of the error on $w_0$ to 1\% by adding a diagonal baryon prior Fisher element, $\alpha F_{bb}$, so that
\begin{equation}
F_{bb}^{\prime} = F_{bb}\left(1+\alpha\right),
\end{equation} 
where $\alpha$ is the external improvement factor. For simplicity we impose a single prior from a broad prospective information increase on baryonic phenomena. We show in Figure ~\ref{fig:fisher_improvement} that reducing degradation to 1\% requires $\alpha=0.47$. This corresponds to an external prior of $\sigma^{\prime}_{b,\rm{con}} = 0.82\,\sigma_{b,\rm{con}}$ in terms of the conditional baryon errors. 
\\
\indent At first, baryon degradation is highly sensitive to relatively small increases in information. This is cause for tentative optimism that if sufficient external data can be used to further inform Stage IV surveys, substantial improvements can be made comparatively easily. However, the rate of improvement with increasing information is soon damped, so significantly stronger priors are required to achieve negligible degradation. 

\begin{table*}
\centering
\begin{tabular}{c|c|c|c|c|c|c}
  \hline
  & $\rm{FOM}_{\rm{WL}}$  & $\rm{FOM}_{\rm{WL}}$ & $R_{\rm{FOM,WL}}$ & $\rm{FOM}_{\rm{WL+CMB}}$  & $\rm{FOM}_{\rm{WL+CMB}}$ & $R_{\rm{FOM,WL+CMB}}$ \\
& (no baryon marg.) & (inc. baryon marg.) & & (no baryon marg.) & (inc. baryon marg.) &   \\
\hline
$w_0$-$w_a$ & 106 & 62.4 & 1.70 & 283 & 145 & 1.96    \\
$n_s$-$\sigma_8$ & 8540 & 1830 & 4.65 & 128000 & 75800 & 1.69 \\
\hline
\end{tabular}
\caption{Figures of merit for $w_0$-$w_a$ and $n_s$-$\sigma_8$ without and including marginalization over baryonic physics, and with and without the addition of priors on $\Lambda$CDM cosmological parameters from \textit{Planck} CMB measurements. For the cases with and without priors, we include the reduction factor, $R_{\rm{FOM}}$, of the FOM when including baryon marginalization to the FOM when baryons are fixed.}
\label{table:priorstable}
\end{table*}

\begin{figure}
\includegraphics[width=\columnwidth]{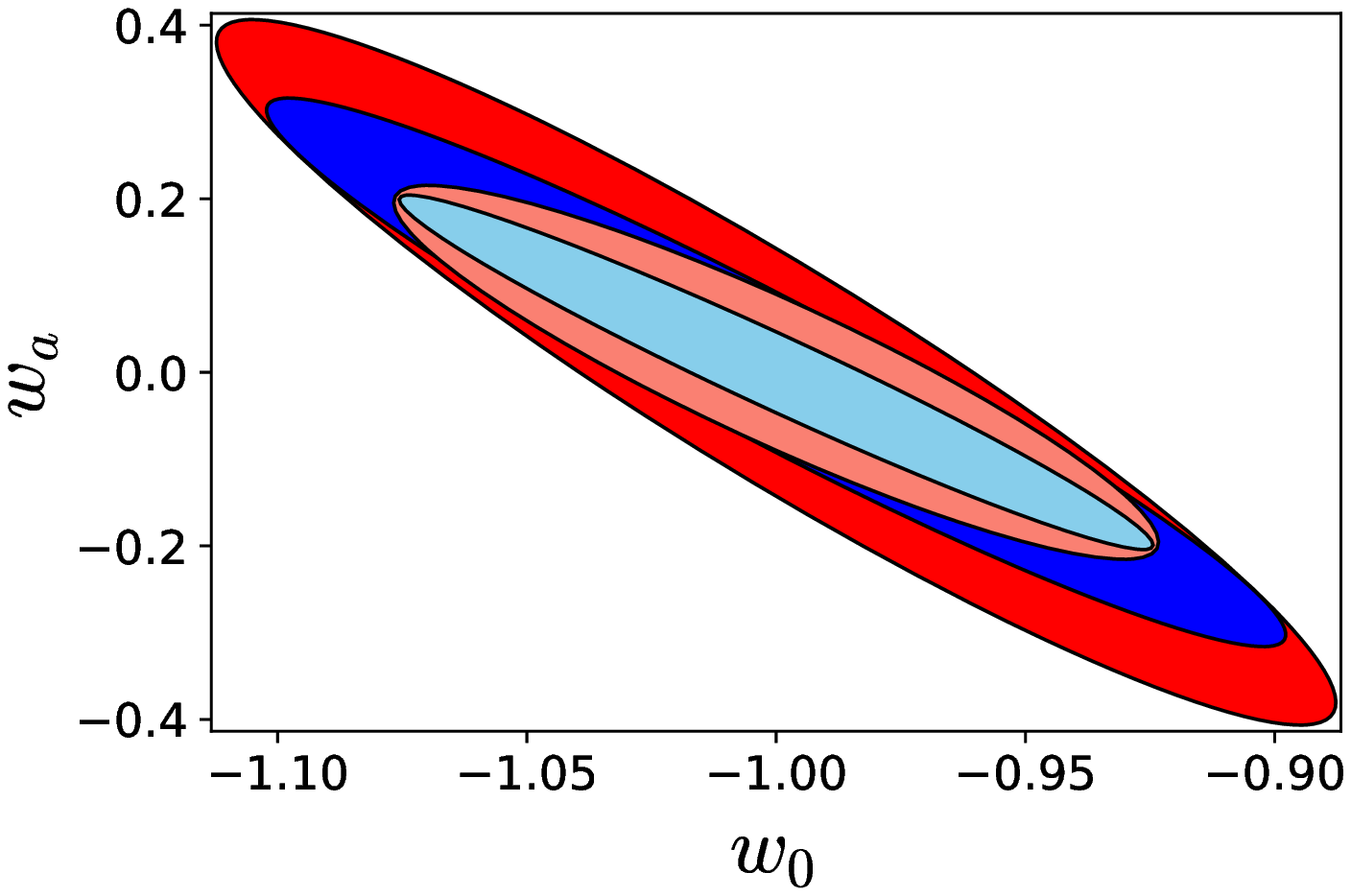}
\caption{1-$\sigma$ 2-parameter confidence ellipses for $w_0$ and $w_a$. In each case, $\Omega_m,\Omega_b,h,n_s,\sigma_8$ have been marginalized over. We show results when all baryon parameters are fixed to their fiducial values (without \textit{Planck} CMB priors: blue; with priors: light blue) and when all baryon parameters are marginalized over (without \textit{Planck} CMB priors: red; with priors: pink).}
\label{fig:lens_w0_wa_istnorm_zbin_priors} 
\end{figure}
\begin{figure}
\includegraphics[width=\columnwidth]{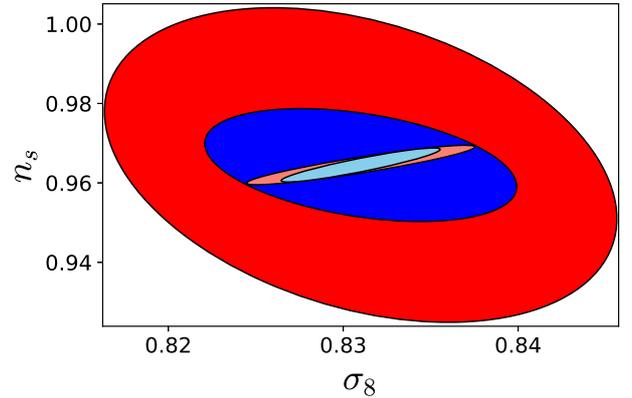}
\caption{1-$\sigma$ 2-parameter confidence ellipses for $\sigma_8$ and $n_s$. In each case, $\Omega_m,\Omega_b,h,w_0,w_a$ have been marginalized over. We show results when all baryon parameters are fixed to their fiducial values (without \textit{Planck} CMB priors: blue; with priors: light blue) and when all baryon parameters are marginalized over (without \textit{Planck} CMB priors: red; with priors: pink).}
\label{fig:lens_sig8_ns_omm_h_planck_eqzbin_priors} 
\end{figure}
\vspace{-2.2mm}

\subsection{Planck CMB Priors}
If the price of external baryon information is too steep, priors can also be added on the cosmological parameters from sources like the early Universe that are independent of Stage IV large-scale structure survey constraints. Inverting the Fisher matrix propagates this information through to the dark energy errors, potentially mitigating baryon degradation. 

Excellent information on the cosmic geometry and matter-energy density is provided by the most recent CMB anisotropy measurements from \textit{Planck}. We use the publicly available MCMC chains for the base $\Lambda$CDM combined TT, TE and EE power spectra \citep[see Table 4 in][]{Planck15}. Constraints from the CMB on $w_0$ and $w_a$ alone are extremely weak without adding information from weak lensing and external sources like BAOs \citep{Planck15XVI}. We therefore use $\Lambda$CDM constraints, which we derive by constructing a covariance matrix from the MCMC chains for $\left(\Omega_m,\Omega_b,h,n_s,\sigma_8\right)$. Inverting this incorporates uncertainties from the cosmological parameters into the resulting prior Fisher matrix, $F_{\rm{CMB}}$ \footnote{Even though the parameter space is non-Gaussian, this approach is consistent within the Fisher approximation.}. The total Fisher information from weak lensing (WL) via a Euclid-like survey and from the CMB via \textit{Planck} is then
\begin{equation}
F_{\rm{tot}} = F_{\rm{WL}} + F_{\rm{CMB}}.
\end{equation}   
Rows and columns of zeroes corresponding to baryon and dark energy parameters have been added to $F_{\rm{CMB}}$ to satisfy the parameter space dimensionality. The resulting improvements on dark energy constraints for $w_0$ and $w_a$ are shown in Figure~\ref{fig:lens_w0_wa_istnorm_zbin_priors}, and for $n_s$ and $\sigma_8$ in Figure~\ref{fig:lens_sig8_ns_omm_h_planck_eqzbin_priors}. Both sets of results are summarized in Table~\ref{table:priorstable}. 

The CMB provides very strong constraints for $n_s$ and $\sigma_8$, dramatically improving the forecast obtained from weak lensing alone, and removing much of the relative baryon degradation. There is an interesting comparison with the $w_0$-$w_a$ constraints. These parameters are not themselves constrained by the CMB but adding the priors still more than doubles the FOM, including when baryons are marginalized over. This is mainly due to breaking degeneracies between dark energy and $\Omega_b$, $h$, and $n_s$. The scale of the transition to non-linear power is affected by $\Omega_b$ and $h$, while $n_s$ tilts $P\left(k\right)$ around $k=1\, h\, \mathrm{Mpc}^{-1}$. Hence, as illustrated in Figures~\ref{fig:multipow_matter_z0} and~\ref{fig:lensing_power_response}, comparable boosts to the non-linear power spectrum occur from raising or lowering these parameters. Dark energy similarly amplifies non-linear power with scale. Adding CMB information on $\Omega_b$, $h$ and $n_s$ alleviates these degeneracies, so the dark energy constraints improve substantially.  

Figure~\ref{fig:lens_w0_wa_istnorm_zbin_priors} shows that one linear combination of ($w_0$, $w_a$) no longer suffers from baryon degradation when CMB priors are included. However there is comparatively limited alleviation of the relative degradation for other directions in the parameter space. This could be due to CMB data being unable to provide information about the relationship between $w_0$ and $w_a$, and baryons. This contributes to the relative degradation of the FOM being largely unchanged after the inclusion of priors. Therefore it should be emphasized that, despite the FOM doubling, the key statistic for constraining the dark energy parameter space is no less impacted by baryons.
\begin{figure}
\includegraphics[width=\columnwidth]{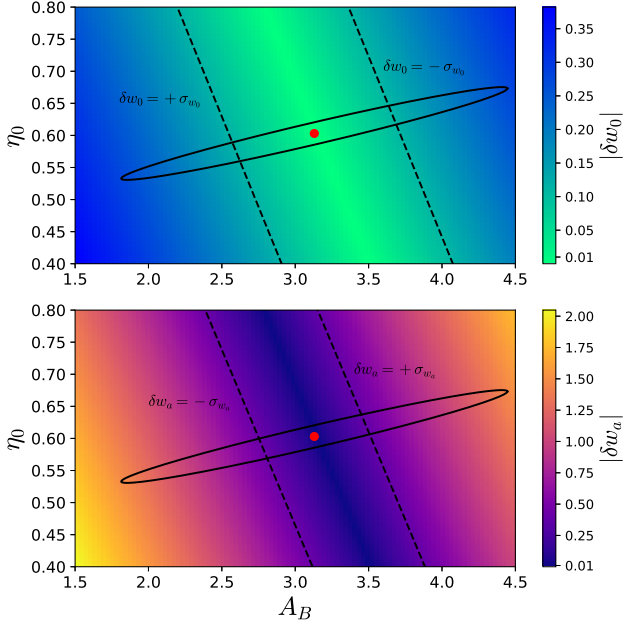}
\caption{Absolute bias in $w_0$ (top) and $w_a$ (bottom) due to model bias in $A_B$ and $\eta_0$. The ellipse represents the marginalized 1-$\sigma$ confidence region for the baryon parameters, with the red dot marking the fiducial point $\left(A_{B,\rm{fid}}=3.13,\eta_{0,\rm{fid}}=0.603\right)$. The dashed lines mark the $A_B$-$\eta_0$ bias corresponding to the marginalized 1-$\sigma$ errors for $w_0$ and $w_a$.}
\label{fig:bias_colourmap} 
\end{figure}

\vspace{-2.2mm}
\section{Model Bias}
\label{sec:modbias}
The fiducial values of $A_B$ and $\eta_0$ have been determined by fitting to simulations in \citetalias{Mead15}. This does not account for systematic limitations of the simulations or incorrect physics. It is important to know how far from the fitted values the true values can lie before $w_0$ and $w_a$ estimates are severely biased. \citet{Taylor07} showed that a first-order approximation of the bias in a cosmological parameter, $\theta$, can be related to the bias in a nuisance parameter, $\psi$, (in our case baryons) through sub-blocks of the full Fisher matrix, such that 
\begin{equation}
\delta\theta_i = -\left[F^{\theta\theta}\right]_{ik}^{-1} F_{kj}^{\theta\psi} \delta\psi_j,
\end{equation}
in which $k$ is implicitly summed over. In Figure~\ref{fig:bias_colourmap} we show the relative biases induced in $w_0$ and $w_a$ when the `true' values of the baryon parameters deviate from the fitted values. Biases of up to $25$\% can occur for $w_0$ if the true values lie at the edges of the ranges $2<A_B<4$ and $0.5<\eta_0<0.8$ found by \citetalias{Mead15} fits to OWLS simulations. A line of minimal bias emerges for both $w_0$ and $w_a$ due to the first-order cancellation of $A_B$ and $\eta_0$ biases. This happens to be almost perpendicular to the minor axis of the marginalized baryon confidence ellipse. 

The $A_B$-$\eta_0$ regions in which the resulting bias to $w_0$ and $w_a$ is within the 1-$\sigma$ marginalized error limits is given by
\begin{equation}
    \left\{
                \begin{array}{ll}
                  |\delta w_0| \leq \sigma_{w_0}, \quad -0.563 A_B + 2.04 \leq \eta_0 \leq -0.563 A_B + 2.69\\
                  |\delta w_a| \leq \sigma_{w_a}, \quad -0.547 A_B + 2.11 \leq \eta_0 \leq -0.547 A_B + 2.53 & .
                \end{array}
              \right.
\end{equation}
A significant proportion of the baryon 1-$\sigma$ confidence region generates an acceptable level of bias. However, positions in the parameter space that would generate biases approaching 35\% for $w_0$, and even more severe effects for $w_a$, remain within the bounds of our forecasts. 

Model bias is difficult to mitigate because it arises directly from subgrid limitations. A full solution likely requires external data on baryon phenomenology, so it is beyond the scope of this work to do more than assess the potential impact of the issue. However, various studies have explored ways to alleviate the issue. \citet{Semboloni13} show that higher order statistics experience different bias due to baryons than two-point shear statistics. They demonstrate that, by combining two- and three-point statistics, the model bias can be mitigated to an extent. Cross-correlations between the thermal Sunyaev-Zeldovich power spectrum and weak lensing observations can also provide valuable information about the baryon distribution directly \citep{Ma15,Hojjati17}. We emphasize that our analysis shows the level of bias that must be overcome by one or more of these methods, and the extent to which results which do not account for it at all can still be considered robust.

\vspace{-2.2mm}

\section{Summary and Conclusions}
\label{sec:conc}
This paper has built upon previous analytic modifications of the halo model to account for the effect of baryonic astrophysical phenomena on the distribution and power spectrum of matter. We used the baryon-halo model of \citetalias{Mead15} to incorporate the impact of adiabatic contraction on halo concentration, and the halo mass-dependent bloating effects of baryonic feedback from e.g., AGN and supernovae. The model of \citetalias{Mead15} was chosen because it provides accurate fits for the power spectrum to within a few percent by calibrating parameters to the COSMIC EMU \citep{Heitmann14} and OWLS \citep{Schaye10} simulations. Other approaches \citep[e.g.,][]{Semboloni11,Mohammed14} focus on precisely modelling stellar, gas and dark matter distributions. The broad, empirically motivated corrections in \citetalias{Mead15} to the power spectrum are instead concerned with the effects of the redistribution of matter under these processes on the halo profile. Our results are therefore underpinned by this approach. We extended the model by incorporating an inner halo core, $r_b$, to account for small-scale structure. This was motivated by an array of baryonic feedback mechanisms, or the condensation of ultra-light axions instead of CDM in the inner halo.

We examined the degradation that marginalizing over the baryon parameters $\left(A_B,\eta_0,r_b\right)$ has on constraints on the $w_0$-$w_a$ dark energy parameter space forecast for a Euclid-like Stage IV cosmological survey. We did this by studying the impact that varying cosmological and baryon parameters has on $P\left(k\right)$ and $C_{\ell}$ at different scales, which informed our interpretation of a full Fisher analysis. The baryon degradation to the errors on $w_0$ and $w_a$ is $\sim$10\% and $\sim$30\% respectively. However, as the FOM is quadratic in parameter uncertainty this translates to a $\sim$40\% degradation in the capacity of a Stage IV survey to deliver accurate measurements. Though we applied our methodology to a Euclid-like survey, it could also be used for other next generation surveys like LSST.

We also highlighted that the effect of baryons is not limited to dark energy, showing the severe degradations on forecasts for $n_s$-$\sigma_8$ errors as an example. This illustrates the potential risk in making confident claims from these surveys even for cosmological parameters which are otherwise well-constrained from sources like the CMB. 

We showed that our framework can forecast constraints on baryons, marginalized over the uncertainty remaining in cosmology. This could potentially provide useful information for modelling baryons in simulations. Euclid-like surveys only constrain $A_B$ and $\eta_0$ at the 50\% and 10\% level respectively, with $\sigma_{A_B}=0.866$ and $\sigma_{\eta_0}=0.0476$, although these improve significantly if the inner core is zero, reducing to $\sigma_{A_B,\rm{cusp}}=0.134$ and $\sigma_{\eta_0,\rm{cusp}}=0.0148$. Our results imply that the inner cores themselves could be constrained to a few $\mathrm{kpc}$. Our implementation of cores is generic so without a more physically motivated treatment these results should not be considered robust. However, this remains an interesting indicator of the capacity of such surveys to potentially forecast inner cores in more sophisticated models. If this result were accurate it would have important implications for the cusp-core debate as it would address the question of cores arising from axion condensation on these scales. 

The degradation we found to dark energy forecasts is of a similar level to \citet{Mohammed14}, who also demonstrate a $\sim$10\% baryon impact on $w_0$ constraints. The model we used has more freedom to vary individual baryonic effects, and has accurate power spectra fits to COSMIC EMU and OWLS. However, the consistent results should be seen as an encouraging sign that the magnitude of the baryonic impact on the $w_0$-$w_a$ parameter space is well-understood. This should temper concerns from the far more pessimistic predictions of \citet{Zentner12} of 50\% level degradations to $w_0$ and $w_a$. The larger impact could be attributed to inaccuracies in the baryon modelling by not accounting for distinct distributions of heated gas and cold dark matter \citep[as noted by][]{Semboloni13}, and calibrating to less accurate power spectra than have since become available \citepalias{Mead15}.

To inform a possible mitigation strategy we first explored the lensing scales on which Fisher information is most sensitive to both cosmological and baryon effects. We found that the region of maximum sensitivity for dark energy occurs at $\ell\sim 100$, i.e. on substantially larger scales than are typically assumed. This is due to the competing effects of geometry and growth broadening the impact of varying $w_0$ and $w_a$ across a wide range of scales. We illustrated that raising $\ell_{\rm{max}}$ has a limited improvement on the FOM and, in fact, suffers from an increasingly worse relative degradation. 

A small amount of external baryon information from simulations or observations provided substantial improvements to the degradation on $w_0$ errors. However, the rate of improvement with information soon tails off so reaching a 1\% degradation threshold requires priors of the order of the baryon conditional errors $\sigma_{\rm{b,prior}} = 0.82 \sigma_{\rm{b,con}}$. This may prove challenging but as significant improvements are still possible, we consider this motivation to acquire stronger observational data for the influence of baryons on large-scale structure.

Constraints on dark energy greatly improved when including the strong cosmological priors offered by \textit{Planck} CMB measurements. Particularly promising was the result that degradation on the errors for one linear combination of $w_0$ and $w_a$ were almost completely removed. An important qualification is that due to dark energy itself being poorly constrained by the CMB there is no relative improvement to the degradation on the cross-covariance between $w_0$-$w_a$. Therefore, while the absolute improvements on the dark energy FOM are significant, the key statistic for constraining the parameter space remains as afflicted by baryons. 

Finally, we considered model bias emerging from incorrect calibrations of baryon parameters. We calculated to first-order the bias that $w_0$ and $w_a$ would experience due to the true values for $A_B$ and $\eta_0$ deviating from fiducial values. Within the $A_B$-$\eta_0$ confidence ellipse there is a significant area corresponding to $w_0$ and $w_a$ bias within 1-$\sigma$ error forecasts. We consider this region to be generally protected from inducing detrimental bias but close to half of the confidence ellipse overlaps with areas of larger biases. It is important to quantify these limitations on our model, though it will ultimately require additional baryon information or improved simulations to fully mitigate the concern.
\par
In summary, by incorporating inner cores into the baryon-halo model of \citetalias{Mead15}, we are able to encompass the full range of broad, empirically motivated baryonic effects on haloes. Our framework allows for quick and flexible predictions on both baryon and dark energy constraints. We anticipate that baryons will have a substantial but not catastrophic effect on the capacity of next generation surveys to constrain dark energy. Mitigation remains an issue. Our thorough examination of the complex interplay of cosmological, baryon and dark energy effects on $C_{\ell}$ showed the limited value of enhancing the survey scope, or redirecting observing power to more linear scales. A combination of external baryon information and CMB priors offers significant improvements and reason for optimism, but there is still work to be done before making degradation negligible.

\vspace{-5.mm}

\section*{Acknowledgements}
We thank John Peacock for a useful discussion, in particular regarding \textit{Planck} CMB priors. We also extend our thanks to an anonymous referee for useful comments. DNC acknowledges the support of an STFC studentship. ANT thanks the Royal Society for a Wolfson Research Merit Award, and the STFC for support from a Consolidated Grant. AH is supported by an STFC Consolidated Grant.




\vspace{-5.mm}

\bibliographystyle{mnras}

\bibliography{The_impact_of_baryons_on_the_sensitivity_of_dark_energy_measurements_version2}

\vspace{-5.mm}

\appendix

\section{Fourier transforms of halo profiles}
\label{appendix:fouriertransforms}
Numerically computing the 1-halo term of the power spectrum given in equation~\eqref{eq:1halo} requires tabulated values of the integrand. This is much less computationally expensive if there is an analytic expression for the window function. This in turn depends on the form of the halo profile in real space that is Fourier transformed. In general,

\begin{equation}
u\left(k\mathopen{|}\mathclose M\right) = \frac{4\pi}{M}\int_0^{r_v}r^2\mathrm{d}r\, \frac{\sin\left(kr\right)}{kr}\rho\left(r, M\right).
\end{equation}
\\
The well-known transform of the NFW profile is given by \citep{Cooray&Sheth02}
\begin{multline}
u_{NFW}\left(k\mathopen{|}\mathclose M_{NFW}\right) = \frac{4\pi\rho_s r_s^3}{M_{NFW}}  \\
 \times \left\{ F\left(k,c\right)\cos\left(kr_s\right) + G\left(k,c\right)\sin\left(kr_s\right) - \frac{\sin\left(ckr_s\right)}{kr_s\left(1+c\right)} \right\},
\end{multline}
where
\begin{align}
F\left(k,c\right) \equiv \mathrm{Ci}\left(\frac{k r_v}{c}\left(1+c\right)\right) - \mathrm{Ci}\left(\frac{k r_v}{c}\right) \nonumber
\\
G\left(k,c\right) \equiv \mathrm{Si}\left(\frac{k r_v}{c}\left(1+c\right)\right) - \mathrm{Si}\left(\frac{k r_v}{c}\right)
\end{align}
and the scale radius, $r_s$, is defined in terms of the concentration, $c$, and virial radius, $r_v$, such that $r_v=cr_s$. Integrating the profile up to the virial radius defines the halo mass,
\begin{equation}
M \equiv 4\pi\int_0^{r_v} r^2 \rho\left(r,M\right)\, \mathrm{d}r,
\end{equation} 
which is evaluated and expressed as a function of the concentration factor,
\begin{equation}
M_{NFW}\left(c\right) = 4\pi\rho_s r_s^3 \left\lbrace \ln\left(1+c\right)-\frac{c}{1+c} \right\rbrace.
\end{equation}
Retaining an analytic Fourier transform when incorporating an inner core is part of the motivation for the simple modification made in equation~\eqref{eq:rhocore}. The resulting window function is
\begin{multline}
u\left(k\mathopen{|}\mathclose M\right) = \frac{4\pi\rho_s r_s^3}{M}\frac{b}{b-c} \Bigl\{ \frac{}{} \frac{M_{NFW}}{4\pi\rho_s r_s^3}u_{NFW}\left(k\mathopen{|}\mathclose M_{NFW}\right) \Bigr. \\
 +\,  \frac{c}{b-c}\frac{1}{k r_s} \left( \left[ \frac{}{} G\left(k,c\right)\cos\left(k r_s\right) - F\left(k,c\right)\sin\left(k r_s\right)\right] \right. \\
\left. \left. \, - \left[G\left(k,b\right)\cos\left(kr_b\right) - F\left(k,b\right)\sin\left(kr_b\right)\frac{}{}\right] \right) \right\},
\end{multline}
\\
where $b=r_v/r_b$ defines an effective `baryon concentration factor'. The halo mass can then be determined as a function of both concentration factors, such that
\begin{multline}
M\left(b,c\right) = \frac{4\pi\rho_s r_s^3}{\left(b-c\right)^2} \left\{ \frac{}{} b\left(b-2c\right)\frac{M_{NFW}}{4\pi\rho_s r_s^3} \right. \\
\left. +\,  c^2 \left[\ln\left(1+b\right) - \frac{b}{1+c}\right] \right\}.
\end{multline}
By taking the limit $r_b\rightarrow{0}$ for both $u$ and $M$, the NFW case is recovered. Despite the simplicity of the $r_b$ modification, it generates a significantly more complex window function. More sophisticated formulations of an inner core typically require numerical Fourier transformations, so they are far less practical for our purposes.

\section{Matter Power Spectrum Responses At Non-Zero Redshifts}
\label{appendix:matterpowz05}
\begin{figure*}
\includegraphics[width=\textwidth]{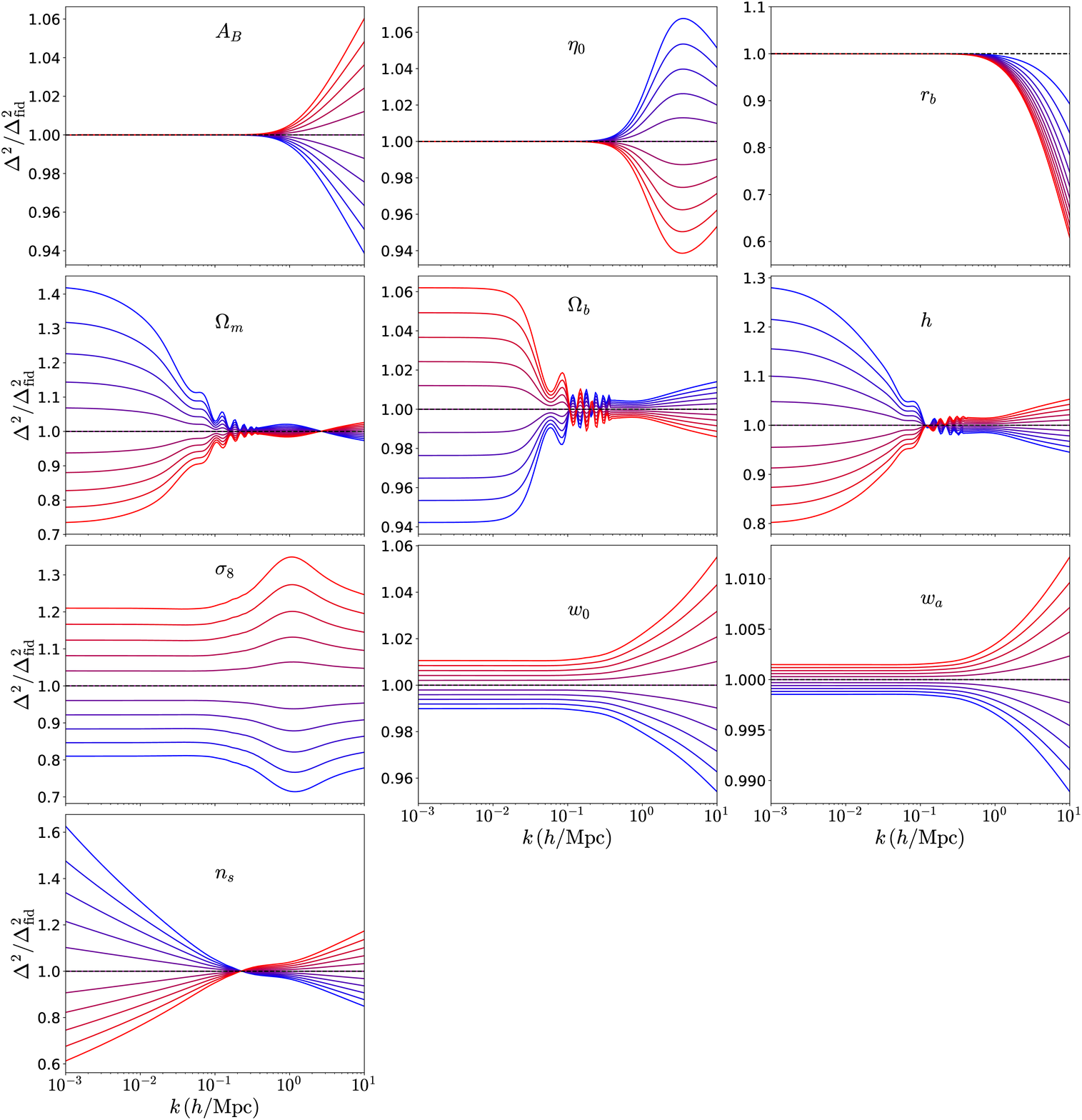}
\caption{The ratio of matter power spectra at $z=0.5$ for different iterations of parameters in 
$\Theta=\left(A_B,\eta_0,\Omega_m,\Omega_b,h,\sigma_8,n_s,w_0,w_a,r_b\right)$, with respect to a fiducial power spectrum given by \textit{Planck} parameters. Bluer (redder) curves correspond to lower (higher) values for parameters in the range $0.9\,\Theta_{\rm{fid}} < \Theta < 1.1\,\Theta_{\rm{fid}}$, except in the case of the dynamic dark energy parameter which varies between $-0.1<w_a<0.1$, and $r_b$ which is varied between core sizes of $r_b=0-100\, h^{-1} \mathrm{kpc}$ and plotted with respect to the fiducial $r_b=0\, h^{-1} \mathrm{kpc}$.}
\label{fig:multipow_matter_z05}
\end{figure*}
In Figure~\ref{fig:multipow_matter_z05} we show the matter power spectrum responses to varying each parameter, at a higher redshift, $z=0.5$, to emphasize the evolution of each influence. Of particular interest are $w_0$ and $w_a$, which uniformly amplify the power on linear scales. This is not the case for $z=0$, where only the non-linear influence of structure affects $P\left(k\right)$. As weak lensing incorporates information from sources and foregrounds over a range of redshifts it is important to be aware of the changes in linear power along the line of sight.

\newpage
\pagebreak
\clearpage
\newpage
\pagebreak
\clearpage
\section{Derivatives of the lensing power spectrum}
\label{appendix:deriv}
\begin{figure}
\includegraphics[width=\columnwidth]{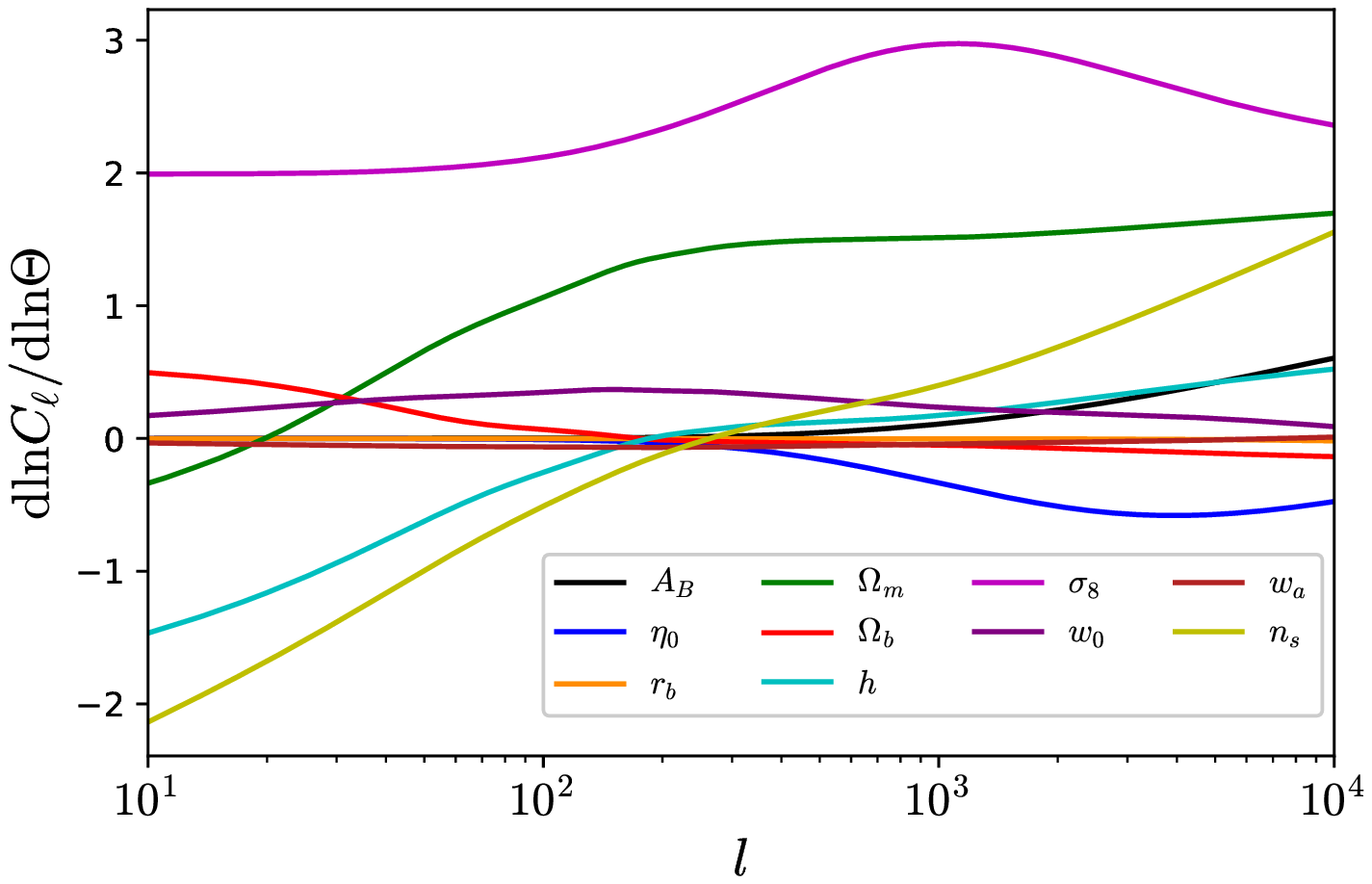}
\caption{Logarithmic derivatives $\mathrm{d}\ln C_{\ell}/\mathrm{d}\theta$ of the weak lensing convergence power spectrum within the redshift bin $0.9 < z < 1.1$, with respect to baryon and cosmological parameters $\Theta=\left(A_B,\eta_0,\Omega_m,\Omega_b,h,\sigma_8,n_s,w_0,w_a,r_b\right)$.}
\label{fig:derivatives}
\end{figure}

The logarithmic derivatives of the weak lensing power spectrum with respect to each parameter are shown in Figure~\ref{fig:derivatives} as this is the essential contribution to the Fisher matrix (somewhat obscured by correlations of power between different redshift bins). We have chosen the redshift bin, $0.9 < z < 1.1$, in which to show the results. We plot the $w_a$ derivative around $w_a=1$ instead of $w_a=0$.

\section{Confidence Ellipses}
\label{appendix:confellip}
\begin{figure*}
\includegraphics[width=\textwidth]{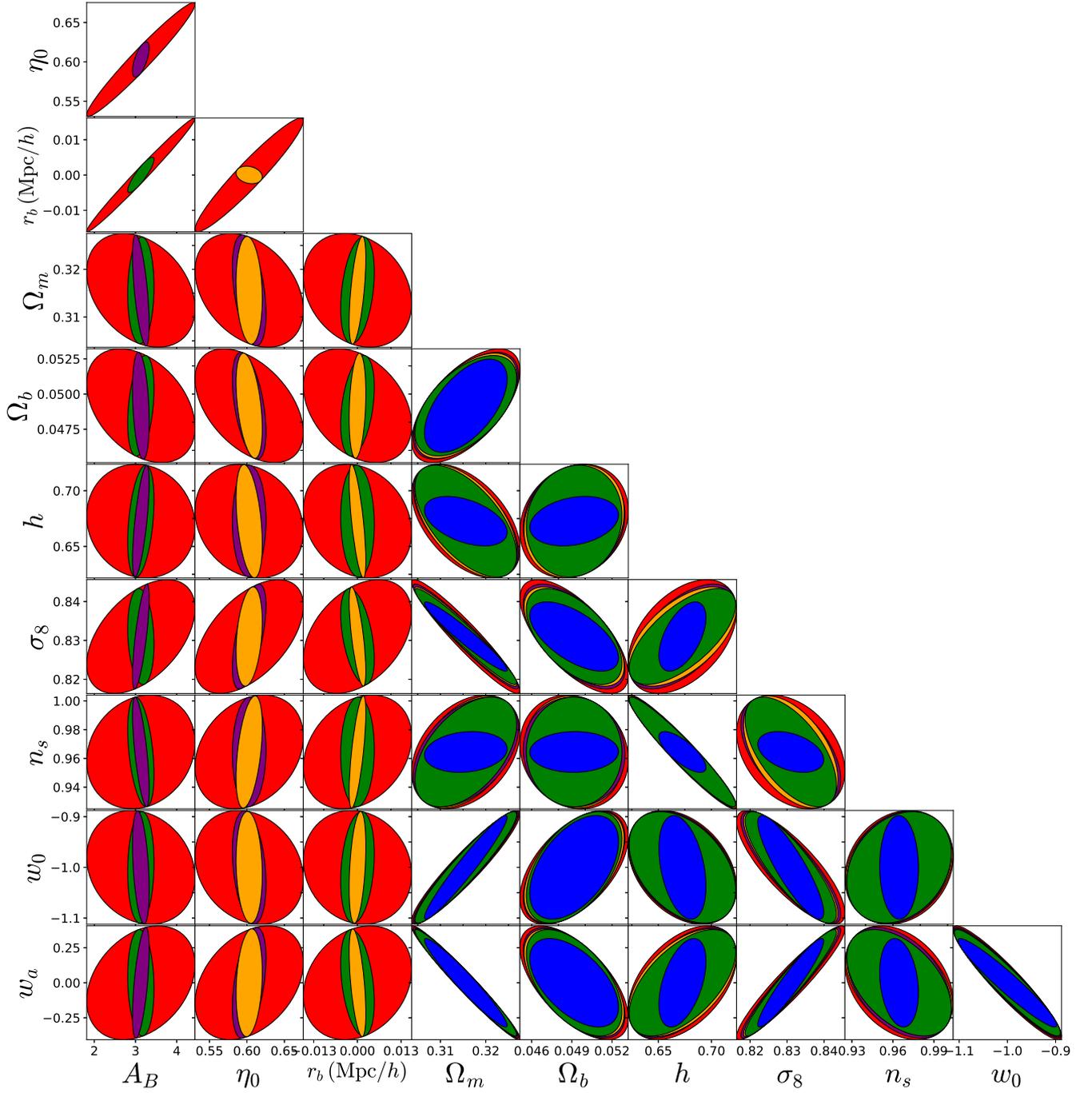}
\caption{1-$\sigma$ 2-parameter confidence ellipses with different combinations of parameters marginalized over: all parameters in $\Theta=\left(A_B,\eta_0,r_b,\Omega_m,\Omega_b, h,\sigma_8,n_s,w_0,w_a\right)$ marginalized over (red); $A_B$ fixed to its fiducial value (orange); $\eta_0$ fixed (green); $r_b$ fixed (purple); $A_B$, $\eta_0$ and $r_b$ fixed (blue).}
\label{fig:fisherlens} 
\end{figure*}

In Figure~\ref{fig:fisherlens} we present the results of the full weak lensing Fisher analysis for every parameter combination in $\Theta=\left(A_B,\eta_0,\Omega_m,\Omega_b,h,\sigma_8,n_s,w_0,w_a,r_b\right)$, with various baryon parameters fixed to their fiducial values or marginalized over.

\bsp 
\label{lastpage}
\end{document}